\documentclass[amsmath,amssymb,aps,prd,reprint,nopreprintnumbers,nofootinbib,showpacs]{revtex4-1}
\usepackage[nolist]{acronym}
\usepackage{multirow}
\usepackage{bigdelim}
\usepackage{multirow}
\usepackage{bm}
\usepackage{graphicx}
\usepackage{mathrsfs}
\usepackage{aas_macros}
\usepackage[colorlinks]{hyperref}
\usepackage{protosem}

\usepackage{listings}
\let\oldmarginpar\marginpar
\renewcommand\marginpar[1]{\-\oldmarginpar[\raggedleft\scriptsize #1]{\raggedright\scriptsize #1}}

\DeclareMathOperator{\Tr}{Tr}
\DeclareMathOperator{\erf}{erf}
\DeclareMathOperator{\cov}{cov}
\DeclareMathOperator{\std}{std}

\DeclareMathOperator*{\argmax}{\arg\!\max}

\def\clap#1{\hbox to 0pt{\hss#1\hss}}

\def\mathrlap{\mathpalette\mathrlapinternal}
\def\mathclap{\mathpalette\mathclapinternal}

\def\mathrlapinternal#1#2{\rlap{$\mathsurround=0pt#1{#2}$}}
\def\mathclapinternal#1#2{\clap{$\mathsurround=0pt#1{#2}$}}

\newcommand\transpose{\ensuremath{^{^\mathsf{T}}}}

\begin{document}

\title[Rapid-response Bayesian sky localization]{Rapid Bayesian position reconstruction for gravitational\nobreakdashes-wave transients}
\author{Leo P. Singer}
\email{leo.singer@ligo.org}
\author{Larry R. Price}
\email{larryp@caltech.edu}
\affiliation{\acs{LIGO} Laboratory, California Institute of Technology, Pasadena, California 91125, USA}
\received{28 September 2015}
\published{14 January 2016}

\begin{abstract}
Within the next few years, Advanced LIGO and Virgo should detect \aclp{GW} from \acl{BNS} and \acl{NSBH} mergers. These sources are also predicted to power a broad array of \acl{EM} transients. Because the \acl{EM} signatures can be faint and fade rapidly, observing them hinges on rapidly inferring the sky location from the \acl{GW} observations. \acl{MCMC} methods for gravitational\nobreakdashes-wave parameter estimation can take hours or more. We introduce BAYESTAR, a rapid, Bayesian, non-\acl{MCMC} sky localization algorithm that takes just seconds to produce probability sky maps that are comparable in accuracy to the full analysis. Prompt localizations from BAYESTAR will make it possible to search \acl{EM} counterparts of compact binary mergers.
\end{abstract}

\maketitle

The Laser Interferometer Gravitational Wave Observatory (\acsu{LIGO})~\cite{AdvancedLIGOHarry,AdvancedLIGO} has just begun taking data \cite{LIGOObservingScenarios} in its ``Advanced'' configuration. The two \ac{LIGO} detectors will ultimately increase their reach in volume within the local Universe by 3 orders of magnitude as compared to their initial configurations through 2010. They form the first parts of a sensitive global gravitational-wave (\acsu{GW}) detector network, soon to be augmented by Advanced Virgo~\cite{AdvancedVirgo} and later by the Japanese KAGRA facility~\cite{LCGT,KAGRAInterferometerDesign} and \ac{LIGO}\nobreakdashes--India~\cite{LIGOIndia}.

The most readily detectable sources of \acp{GW} include binary neutron star mergers, with 0.4\nobreakdashes--400 events per year within the reach of Advanced LIGO at its final design sensitivity~\cite{rates}. These binary systems are not only efficient sources of \acp{GW} but also potential sources of detectable \ac{EM} transients from the aftermath of the tidal disruption of the \acp{NS}. \citet{MostPromisingEMCounterpart} argue that the most promising \ac{EM} counterparts are the hypothesized optical/infrared ``kilonovae'' powered by the radioactive decay of $r$\nobreakdashes-process elements synthesized within the neutron\nobreakdashes-rich ejecta. These are expected to be faint and red and to peak rapidly, reaching an absolute magnitude of only $M_R \sim -13$ within a week, though rising several magnitudes brighter in the infrared~\citep{KilonovaHighOpacities}.

Several mechanisms could make the kilonovae brighter, bluer, and hence more readily detectable \citep{KilonovaPrecursor,KilonovaRedOrBlue}, but peak even earlier, within hours. If, as is widely believed \citep{1986ApJ...308L..43P,1989Natur.340..126E,1992ApJ...395L..83N,2011ApJ...732L...6R}, \ac{BNS} mergers are indeed progenitors of short \acp{GRB}, then a small (due to jet collimation) fraction of Advanced \ac{LIGO} events could also be accompanied by a bright optical afterglow, but this signature, likewise, would peak within hours or faster.

Adding to the challenge of detecting a faint, short-lived optical transient, there is an extreme mismatch between the sky localization accuracy of \ac{GW} detector networks, $\sim$10\nobreakdashes--500\,deg$^2$ \citep{FairhurstTriangulation,WenLocalizationAdvancedLIGO,FairhurstLocalizationAdvancedLIGO,2011PhRvD..84j4020V,RodriguezBasicParameterEstimation,NissankeLocalization,NissankeKasliwalEMCounterparts,KasliwalTwoDetectors,Grover:2013,SiderySkyLocalizationComparison,FirstTwoYears}, and the \acp{FOV} of 1\nobreakdashes--8\,m\nobreakdashes-class optical telescopes. Wide\nobreakdashes-field optical transient facilities such as BlackGEM~(0.6\,m/2.7\,deg$^2$), the \acl{ZTF}~(1.2\,m/47\,deg$^2$)~\cite{ZTF}, the \acl{DECAM}~(4\,m/3\,deg$^2$), or the \acl{LSST}~(8.4\,m/9.6\,deg$^2$), operated in ``target of opportunity'' mode, may be able to tile these large areas rapidly enough to find the one needle in the haystack that is connected to the \ac{GW} event. However, prompt and accurate \ac{GW} position reconstructions will be of the utmost importance for guiding the selection of fields to observe.

The final science run of the initial LIGO and Virgo instruments saw the first joint search for \ac{GW} and \ac{EM} emission from compact binaries. This involved several advances in the \ac{GW} data analysis \citep{CBCLowLatency}, including the first real\nobreakdashes-time matched\nobreakdashes-filter detection pipeline, the \acl{MBTA}~\cite{LIGOVirgoInspiralPipelineComparison}; a semi\nobreakdashes-coherent, \textit{ad hoc} rapid triangulation code, Timing++; and the first version of a rigorous Bayesian \ac{MCMC} parameter estimation code, LALINFERENCE~\cite{S6PE}---all in service of the first search for X\nobreakdashes-ray \citep{SwiftFollowup} and optical \citep{OpticalImageAnalysis} counterparts of \ac{GW} triggers, by a consortium of facilities. Despite the technical achievements in the \ac{GW} data analysis, there was an undesirable tradeoff between the speed as well as accuracy of the rapid localization and the full parameter estimation: the former could analyze a detection candidate in minutes, whereas the latter took half a day; the latter decreased the area on the sky by a factor of 1/20 over the former but took 1000 times as long to run~\cite{SiderySkyLocalizationComparison}.

The success of \ac{EM} follow\nobreakdashes-up of \ac{LIGO} events will depend critically on disseminating high quality sky localizations within a time scale of minutes to hours. To that end, we have devised a rapid and accurate Bayesian sky localization method that takes mere seconds to achieve approximately the same accuracy as the full \ac{MCMC} analysis. Our key insights are the following:
\begin{enumerate}
    \item Nearly all of the information in the \ac{GW} time series that is informative for sky localization is encapsulated within the matched\nobreakdashes-filter estimates of the times, amplitudes, and phases on arrival at the detectors. To infer the position and distance of a \ac{GW} event, we only have to consider three numbers per detector rather than a densely sampled strain time series per detector.
    \item The matched\nobreakdashes-filter pipeline can be treated as a measurement system in and of itself. Just like the strain time series from the detectors, the resultant times, amplitudes, and phases have a predictable and quantifiable measurement uncertainty that can be translated into a likelihood function suitable for Bayesian inference.
    \item The Fisher information matrix will provide clues as to suitable forms of this likelihood function. Recent \ac{GW} parameter estimation literature has largely rejected the Fisher matrix,\footnote{Though not entirely; see \cite{EffectiveFisherMatrix}.} but this is mostly on the grounds of the abuse of the related \ac{CRLB} outside its realm of validity of moderate to high \ac{SNR}~\cite{BayesianBounds,UseAbuseFisherMatrix,FisherMatrixAsymptoticExpansions,InadequaciesFisherMatrix}. However, we recognize that the block structure of the Fisher matrix provides important insights and is a useful quantity for checking the validity of the aforementioned likelihood function, quite independent of the \ac{CRLB}.
    \item The Fisher matrix teaches us that errors in sky localization are semi\nobreakdashes-independent from errors in masses. This implies that if we care only about position reconstruction and not about jointly estimating masses as well, then we can reduce the dimensionality of the parameter estimation problem significantly. Moreover, this frees us of the need to directly compute the expensive post-Newtonian model waveforms, making the likelihood itself much faster to evaluate.
    \item Thanks to a simple likelihood function and a well\nobreakdashes-characterized parameter space, we may dispense with costly and parallelization\nobreakdashes-resistant \ac{MCMC} integration and instead perform the Bayesian marginalization with classic, deterministic, very low order Gaussian quadrature.
    \item The Bayesian inference scheme thus designed to operate on the matched\nobreakdashes-filter detection pipeline outputs could be trivially generalized to operate on the full \ac{GW} time series within the same computational constraints. This would yield a fast and coherent localization algorithm that is mathematically equivalent to the full \ac{MCMC} parameter estimation, restricted to extrinsic parameters (sky location, binary orientation, and distance).
\end{enumerate}
We call this algorithm \ac{BAYESTAR}.\footnote{A pun on the Cylon battleships in the American television series Battlestar Galactica. The defining characteristic of the Cylons is that they repeatedly defeat humanity by using their superhuman information\nobreakdashes-gathering ability to coordinate overwhelming forces.}\footnote{We do not like to mention the final `L' in the acronym, because then it would be pronounced BAYESTARL, which sounds stupid.} It is as fast as Timing++ but nearly as accurate as the rigorous full parameter estimation. It is unique in that it bridges the detection and parameter estimation of \ac{GW} signals, two tasks that have until now involved very different numerical methods and time scales. Beginning with the first Advanced \acs{LIGO} observing run, \ac{BAYESTAR} is providing localizations within minutes of the detection of any \ac{BNS} merger candidate, playing a key role in enabling rapid follow-up observations.

This paper is organized as follows. In Sec.~\ref{sec:detection}, we describe the \ac{GW} signal model and sketch the standard detection algorithm, the matched\nobreakdashes-filter bank. In Sec.~\ref{sec:parameter-estimation}, we describe Bayesian inference formalism and the prevailing method for inferring the parameters of detected candidates, \ac{MCMC} sampling. In Sec.~\ref{sec:bayestar-likelihood}, we propose the \ac{BAYESTAR} likelihood as a model for the uncertainty in the matched\nobreakdashes-filter parameter estimates, and discuss its relationship to and consistency with the likelihood for the full \ac{GW} data. In Sec.~\ref{sec:prior}, we describe the input to \ac{BAYESTAR} supplied by the detection pipeline, and the prior distribution on parameters. In Sec.~\ref{sec:marginal-posterior}, we explain the integration scheme by which the posterior probability distribution is calculated for a given sky location. In Sec.~\ref{sec:adaptive-sampling}, we show a scheme whereby the sky posterior is sampled on an adaptive \ac{HEALPix} grid. In Sec.~\ref{sec:run-time}, we report the running time of the algorithm on the hardware available on the \ac{LIGO} Data Grid. Finally, in Sec.~\ref{sec:case-study}, we quantify the sky localization performance on a comprehensive set of simulated \ac{GW} events.

\section{Signal model and detection}
\label{sec:detection}

In the \acl{TD}, the strain observed by a single \ac{GW} interferometer is
\begin{equation}
    y_i(t) = x_i (t; \bm\theta) + n_i (t).
\end{equation}
In the \acl{FD},
\begin{equation}\label{eq:signal-model}
    Y_i (\omega) = \int_{-\infty}^\infty y(t) e^{-i \omega t} dt = X_i (\omega; \bm\theta) + N_i (\omega),
\end{equation}
where $X_i (\omega; \bm\theta)$ is the \ac{GW} signal given a parameter vector $\bm\theta$ that describes the \ac{GW} source and $N_i (\omega)$ is that detector's Gaussian noise with one\nobreakdashes-sided \ac{PSD} $S_i(\omega) = E\left[\left|N_i(\omega)\right|^2\right] + E\left[\left|N_i(-\omega)\right|^2\right] = 2 E\left[\left|N_i(\omega)\right|^2\right]$. We shall denote the combined observation from a network of detectors as $\mathbf Y (\omega) \equiv \{Y_i (\omega)\}_i$.

Under the assumptions that the detector noise is Gaussian and that the noise from different detectors is uncorrelated, the likelihood of the observation, $\mathbf y$, conditioned upon the parameters $\bm\theta$, is a product of Gaussian distributions:
\begin{multline}\label{eq:gaussian-likelihood}
    \mathcal{L}(\mathbf Y; \bm\theta) = \prod_i p(Y_i | \bm\theta)
        \\
        \propto \exp \left[
        - \frac{1}{2} \sum_i \int_0^\infty \frac{\left|Y_i (\omega)
            - X_i(\omega; \bm\theta) \right|^2}{S_i(\omega)} \, d\omega
    \right].
\end{multline}

A \ac{CBC} source is specified by a vector of extrinsic parameters describing its position and orientation and intrinsic parameters describing the physical properties of the binary components\footnote{This list of parameters involves some simplifying assumptions. Eccentricity is omitted; although it may play a major role in the evolution and waveforms of rare close binaries formed by dynamical capture~\citep{PhysRevD.87.043004,2013PhRvD..87l7501H,2014PhRvD..90h4016H}, \ac{BNS} systems formed by binary stellar evolution should almost always circularize due to tidal interaction~\citep{0004-637X-572-1-407} and later \ac{GW} emission~\citep{PhysRev.136.B1224} long before the inspiral enters LIGO's frequency range of $\sim$10\nobreakdashes--1000~kHz. Tidal deformabilities of the \acp{NS} are omitted because the signal imprinted by the companions' material properties is so small that it will only be detectable by an Einstein Telescope\nobreakdashes-class \ac{GW} observatory~\citep{PhysRevD.81.123016}. Furthermore, in \ac{GW} detection efforts, especially those focused on \ac{BNS} systems, the component spins $\mathbf{S}_1$ and $\mathbf{S}_2$ are often assumed to be nonprecessing and aligned with the system's total angular momentum and condensed to a single scalar parameter $\chi$, or even neglected entirely: $\mathbf{S}_1 = \mathbf{S}_2 = 0$.}:
\begin{equation}\label{eq:params}
    \bm\theta = \left[
    \begin{array}{cl@{\quad}p{4cm}lp{3.75cm}}
        \alpha & \rdelim]{11}{0mm} & right ascension & \rdelim\}{7}{1mm} & \multirow{7}{2cm}{extrinsic parameters, $\bm\theta_\mathrm{ex}$} \\
        \delta && declination & \\
        r && distance & \\
        t_\oplus && arrival time at geocenter & \\
        \iota && inclination angle & \\
        \psi && polarization angle & \\
        \phi_c && coalescence phase & \\
        \cline{1-1}\cline{3-3}
        m_1 && first component's mass & \rdelim\}{4}{1mm} & \multirow{4}{2cm}{intrinsic parameters, $\bm\theta_\mathrm{in}$.}\\
        m_2 && second component's mass & \\
        \mathbf S_1 && first component's spin & \\
        \mathbf S_2 && second component's spin & \\
    \end{array}\right.
\end{equation}

Assuming a non\nobreakdashes-precessing circular orbit, we can write the \ac{GW} signal received by any detector as a linear combination of two basis waveforms, $H_0$ and $H_{\pi/2}$ \citep{PhysRevD.83.084002}. $H_0$ and $H_{\pi/2}$ are approximately ``in quadrature'' in the same sense as the cosine and sine functions, being orthogonal and out of phase by ${\pi/2}$ at all frequencies. If $H_0$ and $H_{\pi/2}$ are Fourier transforms of real functions, then $H_0(\omega) = H_0^*(-\omega)$ and $H_{\pi/2}(\omega) = H_{\pi/2}^*(-\omega)$, and we can write
\begin{equation}
    H_{\pi/2}(\omega) = H_0(\omega) \cdot
    \begin{cases}
        -i & \text{if } \omega \geq 0 \\
        i & \text{if } \omega < 0
    \end{cases}.
\end{equation}
For brevity, we define $H \equiv H_{0}$ and write all subsequent equations in terms of the $H$ basis vector alone. Then, we can write the signal model in a way that isolates all dependence on the extrinsic parameters, $\bm\theta_\mathrm{ex}$, into a few coefficients and all dependence on the intrinsic parameters, $\bm\theta_\mathrm{in}$, into the basis waveform, by taking the Fourier transform of Eq.~(2.8) of Ref.~\cite{PhysRevD.83.084002},
\begin{multline}\label{eq:full-signal-model}
    X_i(\omega; \bm\theta) = e^{-i \omega (t_\oplus - \mathbf{d}_i \cdot \mathbf{n})}
    \frac{r_{1,i}}{r}
    e^{2 i \phi_c}
    \\
    \left[
    \frac{1}{2} \left(1 + \cos^2 \iota\right) \Re \left\{\zeta\right\} - i
    \left(\cos\iota\right) \Im \left\{\zeta\right\}
    \right]
    H(\omega; \bm\theta_\mathrm{in})
\end{multline}
for $\omega \geq 0$, where
\begin{equation}
    \zeta = e^{-2 i \psi} \left(
    F_{+,i}(\alpha, \delta, t_\oplus) +
    i F_{\times,i}(\alpha, \delta, t_\oplus)
    \right).
\end{equation}
The quantities $F_{+,i}$ and $F_{\times,i}$ are the dimensionless detector antenna factors, defined such that $0 \leq {F_{+,i}}^2 + {F_{\times,i}}^2 \leq 1$. They depend on the orientation of detector $i$ as well as the sky location and sidereal time of the event and are presented in Ref.~\cite{ExcessPower}. In a coordinate system with the $x$ and $y$ axes aligned with the arms of a detector, the antenna pattern is given in spherical polar coordinates as
\begin{align}
    F_+ &= -\frac{1}{2}(1 + \cos^2 \theta) \cos{2\phi}, \\
    F_\times &= -\cos\theta \sin{2\phi}.
\end{align}
The unit vector $\mathbf{d}_i$ represents the position of detector $i$ in units of light travel time.\footnote{When considering transient \ac{GW} sources such as those that we are concerned with in this article, the origin of the coordinate system is usually taken to be the geocenter. For long\nobreakdashes-duration signals such as those from statically deformed neutron stars, the solar system barycenter is a more natural choice.} The vector $\mathbf{n}$ is the direction of the source. The negative sign in the dot product $-\mathbf{d}_i \cdot \mathbf{n}$ is present because the direction of travel of the \ac{GW} signal is opposite to that of its sky location. The quantity $r_{1,i}$ is a fiducial distance at which detector $i$ would register \ac{SNR}=1 for an optimally oriented binary (face\nobreakdashes-on, and in a direction perpendicular to the interferometer's arms):
\begin{equation}\label{eq:horizon}
r_{1,i} = 1 / \sigma_i, \qquad {\sigma_i}^2 = \int_0^\infty \frac{\left|H(\omega; \boldsymbol \theta_\mathrm{in})\right|^2}{S_i(\omega)} \,d\omega.
\end{equation}

More succinctly, we can write the signal received by detector $i$ in terms of observable extrinsic parameters $\bm\theta_i = (\rho_i, \gamma_i, \tau_i)$, the amplitude $\rho_i$, phase $\gamma_i$, and time delay $\tau_i$ on arrival at detector $i$:
\begin{multline}\label{eq:signal-model}
    X_i (\omega; \bm\theta_i, \bm\theta_\mathrm{in})
    \\
    = X_i (\omega; \rho_i, \gamma_i, \tau_i, \bm\theta_\mathrm{in}) = \frac{\rho_i}{\sigma_i} e^{i (\gamma_i - \omega \tau_i)} H(\omega; \bm\theta_\mathrm{in}).
\end{multline}

The prevailing technique for detection of \acp{GW} from \acp{CBC} is to realize a \ac{ML} estimator (MLE) from the likelihood in Eq.~(\ref{eq:gaussian-likelihood}) and the signal model in Eq.~(\ref{eq:signal-model}). Concretely, this results in a bank of matched filters, or cross-correlations between the incoming data stream and a collection of template waveforms,
\begin{equation}
z_i(\tau_i;\bm\theta_\mathrm{in}) = \frac{1}{\sigma_i (\bm\theta_\mathrm{in})} \int_0^\infty \frac{H^*(\omega; \bm\theta_\mathrm{in}) Y_i(\omega) e^{i \omega \tau_i}}{S_i(\omega)} \,d\omega.
\end{equation}
The \ac{ML} point estimates of the signal parameters, $\mathrm{MLE}(\mathbf{y}) = \{\{ \hat{\bm\theta}_i \}_i, \hat{\bm\theta}_\mathrm{in}\} = \{\left\{ \hat\rho_i, \hat\gamma_i, \hat\tau_i \right\}_i, \hat{\bm\theta}_\mathrm{in}\}$, are given by
\begin{eqnarray}
    \label{eq:optimal-tau}
    \hat{\bm\theta}_\mathrm{in}, \{\hat\tau_i\}_i
        &=& \argmax_{\bm\theta_\mathrm{in}, \{\hat\tau_i\}_i}
        \sum_i \left| z_i\left(\tau_i;
        \bm\theta_\mathrm{in}\right) \right|^2, \\
    \label{eq:optimal-rho}
    \hat\rho_i &=& \left| z_i\left(\hat\tau_i;
        \hat{\bm\theta}_\mathrm{in}\right) \right|, \\
    \label{eq:optimal-gamma}
    \hat\gamma_i &=& \arg z_i\left(\hat\tau_i;
        \hat{\bm\theta}_\mathrm{in}\right).
\end{eqnarray}
A detection candidate consists of $\{\left\{ \hat\rho_i, \hat\gamma_i, \hat\tau_i \right\}_i, \hat{\bm\theta}_\mathrm{in}\}$. There are various ways to characterize the significance of a detection candidate. In Gaussian noise, the maximum likelihood for the network is obtained by maximizing the network \ac{SNR}, $\rho_\mathrm{net}$,
\begin{equation}
    \hat\rho_\mathrm{net} = \max_{\bm\theta} \sum_i {|z_i({\bm\theta})|}^2 = \sqrt{\sum_i {\hat\rho_i}^2};
\end{equation}
this, therefore, is the simplest useful candidate ranking statistic.

\subsection{Uncertainty and the Fisher matrix}
\label{sec:fisher}

We can predict the uncertainty in the detection pipeline's \ac{ML} estimates using the \ac{CRLB}. The \ac{CRLB} has been widely applied in \ac{GW} data analysis to estimate parameter estimation uncertainty \cite{1996PhRvD..53.3033B,FairhurstTriangulation,2009PhRvD..79h4032A,WenLocalizationAdvancedLIGO,LIGOObservingScenarios,FairhurstLIGOIndia}.\footnote{\label{footnote:template-banks}The Fisher matrix is also used in construction of \ac{CBC} matched-filter banks. The common procedure is to place templates uniformly according to the determinant of the signal space metric, which is the Fisher matrix. This is equivalent to uniformly sampling the Jeffreys prior. In practice, this is done either by constructing a hexagonal lattice~\citep{PhysRevD.76.102004} or sampling stochastically~\citep{2009PhRvD..80j4014H,2009PhRvD..80b4009V,SBank,2010PhRvD..81b4004M,2014PhRvD..89b4003P}.} As we noted, there are significant caveats to the \ac{CRLB} at low or moderate \ac{SNR} \cite{BayesianBounds,UseAbuseFisherMatrix,FisherMatrixAsymptoticExpansions,InadequaciesFisherMatrix}. However, here we will be concerned more with gaining intuition from the block structure of the Fisher matrix than its numerical value. Furthermore, the Fisher matrix in its own right---independent of its suitability to describe the parameter covariance---is a well\nobreakdashes-defined property of any likelihood function, and we will exploit it as such in Sec.~\ref{sec:bayestar-likelihood}.

We will momentarily consider the likelihood for a single detector:
\begin{multline}\label{eq:gaussian-likelihood-spa}
    \mathcal{L}\left(Y_i; \rho_i, \gamma_i, \tau_i,
        \bm\theta_\mathrm{in}\right)
    \\
    \propto \exp \left[
        - \frac{1}{2} \int_0^\infty \frac{\left|Y_i (\omega)
            - X_i\left(\omega; \rho_i, \gamma_i, \tau_i,
                \bm\theta_\mathrm{in}\right)
        \right|^2}{S_i(\omega)} \, d\omega
    \right],
\end{multline}
with $X_i(\omega; \rho_i, \gamma_i, \tau_i, \bm\theta_\mathrm{in})$ given by Eq.~(\ref{eq:signal-model}).

The Fisher information matrix for a measurement $y$ described by the unknown parameter vector $\bm{\theta}$ is the conditional expectation value
\begin{equation}\label{eq:general-fisher-matrix}
    \mathcal{I}_{jk} = \mathrm{E} \, \left[
        \left(\frac{\partial \log
            \mathcal{L}(Y_i ; \bm\theta)}
            {\partial \theta_j}\right)
        \left(\frac{\partial \log
            \mathcal{L}(Y_i ; \bm\theta)}
            {\partial \theta_k}\right)
    \middle| \bm\theta
    \right].
\end{equation}

The Fisher matrix describes how strongly the likelihood depends, on average, on the parameters. Furthermore, it provides an estimate of the mean-square error in the parameters. If $\hat{\bm\theta}$ is an unbiased estimator of $\bm\theta$, $\tilde{\bm\theta} = \hat{\bm\theta} - \bm\theta$ is the measurement error, and $\Sigma = \mathrm{E} \, [\tilde{\bm\theta}\tilde{\bm\theta}\transpose]$ is the covariance of the measurement error, then the \ac{CRLB} says that $\Sigma \geq \mathcal{I}^{-1}$, in the sense that $\left(\Sigma - \mathcal{I}^{-1}\right)$ is positive semidefinite.

Note that if $\log\mathcal{L}$ is twice differentiable in terms of $\bm\theta$, then the Fisher matrix can also be written in terms of second derivatives as
\begin{equation}\label{eq:general-fisher-matrix-second-derivatives}
    \mathcal{I}_{jk} = \mathrm{E} \, \left[
        -\frac{\partial^2 \log
            \mathcal{L}(Y_i ; \bm\theta)}
            {\partial \theta_j \partial \theta_k}
    \middle| \bm\theta
    \right].
\end{equation}

When (as in our assumptions) the likelihood is Gaussian,\footnote{This assumes that the merger occurs at a frequency outside the sensitive ``bucket'' of the detector's noise \ac{PSD}. There are additional terms if the \ac{GW} spectrum drops to zero within the sensitive bandwidth of the detector, as can be the case for \acl{NSBH} mergers; see Ref.~\cite{2014CQGra..31o5005M}.} Eq.~(\ref{eq:general-fisher-matrix}) simplifies to
\begin{equation}\label{eq:gaussian-fisher-matrix}
    \mathcal{I}_{jk} = \int_0^\infty \Re \left[
        \left(\frac{\partial X_i}{\partial \theta_j}\right)^*
        \left(\frac{\partial X_i}{\partial \theta_k}\right)
    \right] \frac{1}{S_i(\omega)} \, d\omega.
\end{equation}
This form is useful because it involves manipulating the signal $X_i (\omega)$ rather than the entire observation $Y (\omega)$. In terms of the $k$th \ac{SNR}-weighted moment of angular frequency,
\begin{equation}\label{eq:angular-frequency-moments}
    {\overline{\omega^k}}_i =
        \left[ \int_0^\infty \frac{|h (\omega)|^2}{S_i(\omega)} \omega^k \, d\omega \right]
        \left[ \int_0^\infty \frac{|h (\omega)|^2}{S_i(\omega)} \, d\omega \right]^{-1},
\end{equation}
the Fisher matrix for the signal in the $i$th detector is
\begin{equation}
    \label{eq:single-detector-block-fisher-matrix}
    \mathcal{I}_i = \left(
        \begin{array}{cc}
            \mathcal{I}_{\bm\theta_i,\bm\theta_i} &
            \mathcal{I}_{\bm\theta_i,\bm\theta_\mathrm{in}} \\
            {\mathcal{I}_{\bm\theta_i,\bm\theta_\mathrm{in}}^{^\mathsf{T}}} &
            {\rho_i}^2 \mathcal{I}_{\bm\theta_\mathrm{in},\bm\theta_\mathrm{in}}
        \end{array}
    \right).
\end{equation}
The top-left block describes only the extrinsic parameters, and is given by
\begin{equation}\label{eq:fisher-matrix}
    \mathcal{I}_{\bm\theta_i,\bm\theta_i} = \bordermatrix{
        ~ & \rho_i & \gamma_i & \tau_i \cr
        \rho_i & 1 & 0 & 0 \cr
        \gamma_i & 0 & {\rho_i}^2 & -{\rho_i}^2 {\overline{\omega}}_i \cr
        \tau_i & 0 & -{\rho_i}^2 {\overline{\omega}}_i & {\rho_i}^2 {\overline{\omega^2}}_i
    }.
\end{equation}
(This is equivalent to an expression given in Ref.~\cite{Grover:2013}.) The bottom row and right column of Eq.~(\ref{eq:single-detector-block-fisher-matrix}) describe the intrinsic parameters and how they are coupled to the extrinsic parameters. We show in Appendix~\ref{sec:independence-intrinsic-extrinsic} that we need not consider the intrinsic parameters at all if we are concerned only with sky localization.

For our likelihood, the \ac{CRLB} implies that
\begin{equation}\label{eq:covariance-matrix}
    \cov{
        \left(
        \begin{array}{c}
            \tilde{\rho}_i \\
            \tilde{\gamma}_i \\
            \tilde{\tau}_i
        \end{array}
        \right)
    } \geq \mathcal{I}^{-1} = \left(
        \begin{array}{ccc}
            1 & 0 & 0 \\
            0 & {\rho_i}^2 {\overline{\omega^2}}_i/{\omega_{\mathrm{rms},i}}^2 & {\rho_i}^2 {\overline{\omega}}_i/{\omega_{\mathrm{rms},i}}^2 \\
            0 & {\rho_i}^2 {\overline{\omega}}_i/{\omega_{\mathrm{rms},i}}^2 & {\rho_i}^2/{\omega_{\mathrm{rms},i}}^2
        \end{array}
    \right),
\end{equation}
where ${\omega_{\mathrm{rms},i}}^2 = {\overline{\omega^2}}_i - {{\overline{\omega}}_i}^2$. This structure implies that errors in the \ac{SNR} are uncorrelated with errors in time and phase and that there is a particular sum and difference of the times and phases that are measured independently (see Appendix~\ref{sec:interpretation-of-errors}).

Reading off the $\tau \tau$ element of the covariance matrix reproduces the timing accuracy in Eq.~(24) of Ref.~\cite{FairhurstTriangulation},
\begin{equation}\label{eq:timing-crlb}
    \std \left(\hat{\tau}_i - \tau_i \right) \geq \sqrt{\left(\mathcal{I}^{-1}\right)_{\tau\tau}} = \frac{\rho_i}{\omega_{\mathrm{rms},i}}.
\end{equation}
\citet{FairhurstTriangulation} goes on to frame the characteristic position reconstruction accuracy of a \ac{GW} detector network in terms of time delay triangulation, with the above formula describing the time of arrival uncertainty for each detector. In Appendix~\ref{sec:position-resolution}, we show how to extend this formalism to include the phases and amplitudes on arrival as well.

\section{Bayesian probability and parameter estimation}
\label{sec:parameter-estimation}

In the Bayesian framework, parameters are inferred from the data by forming the posterior distribution, $p(\bm\theta|\mathbf y)$, which describes the probability of the parameters given the observations. Bayes's rule relates the likelihood $p(\mathbf y|\bm\theta)$ to the posterior $p(\bm\theta|\mathbf y)$,
\begin{equation}\label{bayes}
p(\bm\theta|\mathbf y) = \frac{p(\mathbf y|\bm\theta) p(\bm\theta)}{p(\mathbf y)},
\end{equation}
introducing the prior distribution $p(\bm\theta)$ which encapsulates previous information about the parameters (for example, arising from earlier observations or from known physical bounds on the parameters) and the evidence $p(\mathbf y)$ which can be thought of as a normalization factor or as describing the parsimoniousness of the model.

The choice of prior is determined by one's astrophysical assumptions. During \acl{S6} when LIGO's Bayesian \ac{CBC} parameter estimation pipelines were first deployed, the prior was taken to be isotropic in the sky location and binary orientation and uniform in volume, arrival time, and the component masses~\cite{S6PE}.

In Bayesian inference, although it is often easy to write down the likelihood or even the full posterior in closed form, usually one is interested in only a subset $\bm\beta$ of all of the model's parameters, the others $\bm\lambda$ being nuisance parameters. In this case, we integrate away the nuisance parameters, forming the marginal posterior
\begin{equation}\label{eq:marginal-posterior}
    p(\bm\beta|\mathbf y) = \int \frac{p(\mathbf y|\bm\beta,\bm\lambda) p(\bm\beta,\bm\lambda)}{p(\mathbf y)} \,d\bm\lambda
\end{equation}
with $\bm\theta = (\bm\beta, \bm\lambda)$. For instance, for the purpose of locating a \ac{GW} source on the sky, all parameters but $(\alpha, \delta)$ are nuisance parameters.

\section{\acs{BAYESTAR} likelihood}
\label{sec:bayestar-likelihood}

For the purpose of rapid sky localization, we assume that we do not have access to the \ac{GW} data $\mathbf{Y}$ itself and that our only contact with it is through the \ac{ML} parameter estimates $\{\left\{ \hat\rho_i, \hat\gamma_i, \hat\tau_i \right\}_i, \hat{\bm\theta}_\mathrm{in}\}$. Although this is a significant departure from conventional \ac{GW} parameter estimation techniques, we can still apply the full Bayesian machinery of Eq.~(\ref{eq:marginal-posterior}) to compute a posterior distribution for the sky location.

The relevant likelihood is now the probability of the \ac{ML} estimates, conditioned upon the true parameter values, and marginalized over all possible \ac{GW} observations:
\begin{equation}\label{eq:detection-candidate-likelihood}
    p\left(\{\hat{\bm\theta}_i\}_i,
        \hat{\bm\theta}_\mathrm{in}
    \middle| \bm\theta\right)
    \propto \int\limits_{\mathclap{\mathbf{Y} | \mathrm{MLE}(\mathbf{Y}) =
        \{\{\hat{\bm\theta}_i\}_i,
        \hat{\bm\theta}_\mathrm{in}\}}}
    p(\mathbf{Y} | \bm\theta) \, p(\bm\theta)
    \, d\mathbf{Y}.
\end{equation}
Although we may not be able to evaluate this equation directly, with some educated guesses we can create a likelihood that has many properties in common with it. Any valid approximate likelihood must have the same Fisher matrix as shown in Eq.~(\ref{eq:single-detector-block-fisher-matrix}). It must also have the same limiting behavior: it should be periodic in the phase error $\tilde{\gamma}_i$ and go to zero as $\tilde{\tau}_i \rightarrow \pm \infty$, $\hat{\rho}_i \rightarrow 0$, or $\hat{\rho}_i \rightarrow \infty$. Additionally, when $\tilde{\tau}_i = 0$, the distribution of ${\hat{\rho}_i}^2$ should reduce to a noncentral $\chi^2$ distribution with 2 degrees of freedom, centered about ${\rho_i}^2$, because the complex matched-filter time series $z_i(t)$ is Gaussian (under the ideal assumption the \ac{GW} strain time series is Gaussian).

These conditions could be satisfied by realizing a multivariate Gaussian distribution with covariance matrix $\Sigma = \mathcal{I}^\intercal$ and then replacing individual quadratic terms in the exponent of the form $-\tilde{\theta}^2/2$ with $\cos{\tilde{\theta}}$.

\begin{figure*}
    \includegraphics[width=\textwidth]{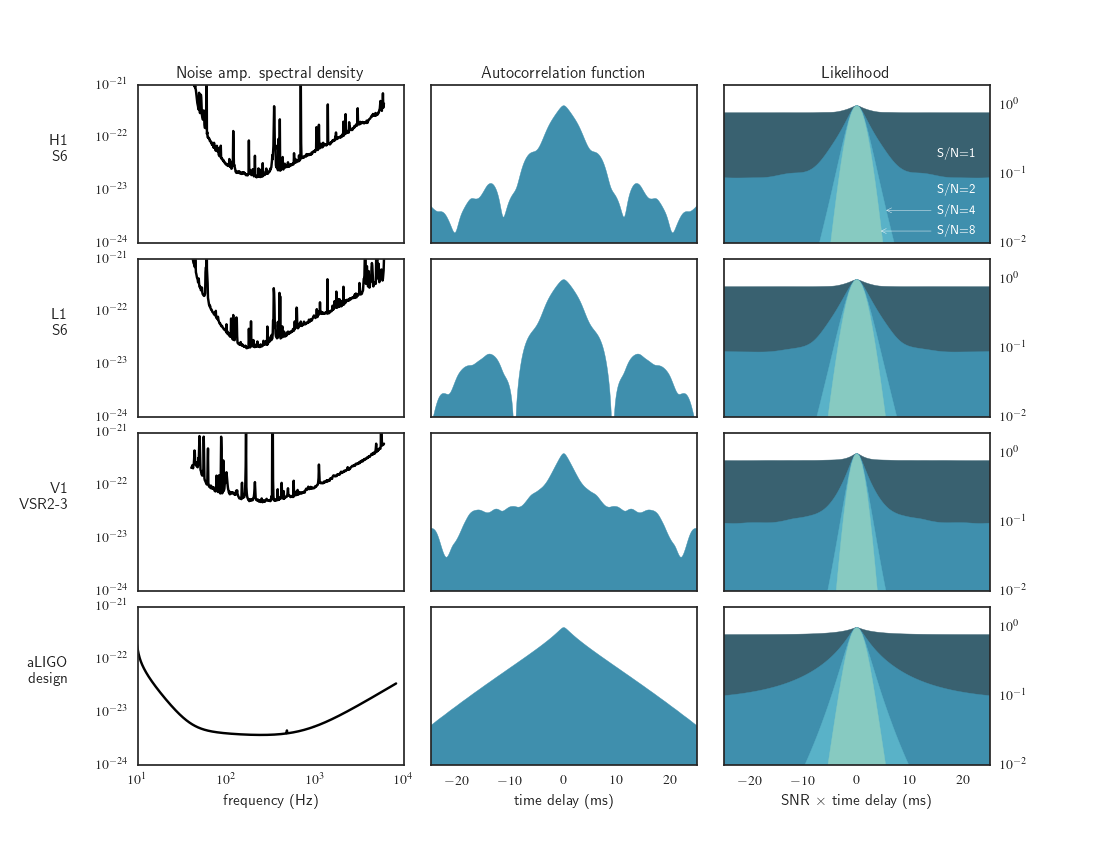}
    \caption{\label{fig:autocorr-likelihood}The autocorrelation likelihood for a $(1.4, 1.4)$\,$M_\odot$ binary as observed by four detector configurations: from top to bottom, the final sensitivity achieved by the \acs{LIGO} Hanford, \acs{LIGO} Livingston, and Virgo detectors in their ``initial'' configuration and the final Advanced \acs{LIGO} design sensitivity. The left panels show the noise amplitude spectral densities. The middle panels show the absolute value of the autocorrelation function. The right panel shows the phase-marginalized autocorrelation likelihood for \ac{SNR}=1,~2,~4,~and~8. In the right panel, the time scale is normalized by the \ac{SNR} so that one can see that as the \ac{SNR} increases, a central parabola is approached (the logarithm of a Gaussian distribution with standard deviation given by the Fisher matrix).}
\end{figure*}

A more natural way is to plug the signal model from Eq.~(\ref{eq:signal-model}) evaluated at the \ac{ML} parameter estimates into the single-detector likelihood in Eq.~(\ref{eq:gaussian-likelihood-spa}):
\begin{multline}
    p\left(\hat{\bm\theta}_i \middle| \bm\theta \right)
    :=
    p\left(Y_i(\omega) = X_i(\omega; \hat{\bm\theta})
        \middle| \bm\theta \right)
    \\
    \propto \exp \Bigg[
        - \frac{1}{2} \int_0^\infty \Bigg|
            \frac{\hat{\rho}_i}{\sigma_i(\hat{\bm\theta}_\mathrm{in})} e^{i (\hat\gamma_i - \omega \hat\tau_i)} \frac{H(\omega; \hat{\bm\theta}_\mathrm{in})}{S_i(\omega)}
    \\
            - \frac{\rho_i}{\sigma_i(\bm\theta_\mathrm{in})} e^{i (\gamma_i - \omega \tau_i)} \frac{H(\omega; \bm\theta_\mathrm{in})}{S_i(\omega)}
        \Bigg|^2 \, d\omega
    \Bigg].
\end{multline}
If we further assume that the intrinsic parameters are equal to their \ac{ML} estimates, $\bm\theta_\mathrm{in} = \hat{\bm\theta}_\mathrm{in}$, then this reduces to what we call the autocorrelation likelihood,
\begin{multline}\label{eq:autocor-likelihood}
    p\left(\hat\rho_i, \hat\gamma_i, \hat\tau_i
        \middle| \rho_i, \gamma_i, \tau_i \right)
    \\
    \propto
    \exp \left[ - \frac{1}{2}{\hat\rho_i}^2 - \frac{1}{2}{\rho_i}^2
        + \hat\rho_i \rho_i \Re \left\{ e^{i \tilde{\gamma}_i} a_i^*(\tilde{\tau}_i)
        \right\}
    \right],
\end{multline}
with $\tilde{\gamma}_i = \hat\gamma_i - \gamma_i$, $\tilde{\tau}_i = \hat\tau_i - \tau_i$, and the template's autocorrelation function $a_i(t; \bm\theta_\mathrm{in})$ defined as
\begin{equation}\label{eq:autocorrelation-function}
    a_i(t; \bm\theta_\mathrm{in}) := \frac{1}{{{\sigma_i}^2(\bm\theta_\mathrm{in})}} \int_0^\infty \frac{\left| H(\omega; \hat{\bm\theta}_\mathrm{in})\right|^2}{S_i(\omega)} e^{i \omega t} \,d\omega.
\end{equation}
Some example autocorrelation functions and corresponding likelihoods are shown in Fig.~\ref{fig:autocorr-likelihood}. To assemble the joint likelihood for the whole network, we form the product of the autocorrelation likelihoods from the individual detectors:
\begin{multline}
    p\left(\left\{\hat\rho_i, \hat\gamma_i, \hat\tau_i\right\}_i
        \middle| \left\{\rho_i, \gamma_i, \tau_i\right\}_i \right) \\
    \propto
    \exp \left[ - \frac{1}{2} \sum_i {\hat\rho_i}^2 - \frac{1}{2} \sum_i {\rho_i}^2
        + \sum_i \hat\rho_i \rho_i \Re \left\{ e^{i \tilde{\gamma}_i} a^*(\tilde{\tau}_i)
        \right\}
    \right].
\end{multline}

In the following section, we discuss some key properties of the autocorrelation likelihood.

\subsection{Properties}
\label{sec:properties}

First, the autocorrelation likelihood has the elegant feature that if we were to replace the autocorrelation function with the \ac{SNR} time series for the best-matching template, $z(\tau; \hat{\mathbf{\theta}}_\mathrm{in})$, we would recover the likelihood for the full \ac{GW} time series, evaluated at the \ac{ML} estimate of the intrinsic parameters, viz.:
\begin{equation}\label{eq:extrinsic-only-likelihood}
                \exp \left[ - \frac{1}{2} \sum_i {\rho_i}^2
        + \sum_i \rho_i \Re \left\{ e^{-i \gamma_i} z_i^*(\tau_i)
        \right\}
    \right].
\end{equation}
[We have omitted the term $\int |Y_i(\omega)|^2/S(\omega)d\omega$, which takes the place of the earlier ${\hat{\rho}_i}^2$ term and is only important for normalization.] The numerical scheme that we will develop is thus equally applicable for rapid, coincidence-based localization, or as a fast extrinsic marginalization step for the full parameter estimation.

Second, observe that at the true parameter values, $\hat{\bm\theta}_i = \bm\theta_i$, the logarithms of Eqs.~(\ref{eq:autocor-likelihood}) and~(\ref{eq:gaussian-likelihood-spa}) have the same Jacobian. This is because the derivatives of the autocorrelation function are
\begin{equation*}
    a^{(n)}(t) = i^n \overline{\omega^n},
\end{equation*}
with $\overline{\omega^n}$ defined in Eq.~(\ref{eq:angular-frequency-moments}). For example, the first few derivatives are
\begin{equation*}
    a(0) = 1,
    \qquad
    \dot{a}(0) = i \overline{\omega},
    \qquad
    \ddot{a}(0) = -\overline{\omega^2}.
\end{equation*}

Using Eq.~(\ref{eq:general-fisher-matrix-second-derivatives}), we can compute the Fisher matrix elements for the autocorrelation likelihood given by Eq.~(\ref{eq:autocor-likelihood}), with the detector subscript suppressed,
\begin{align}
    \nonumber
    \mathcal{I}_{\rho\rho} &= 1, \\
    \nonumber
    \mathcal{I}_{\rho\gamma} &= 0, \\
    \nonumber
    \mathcal{I}_{\rho\tau} &= 0, \\
    \label{eq:fisher-autocor-gamma-gamma}
    \mathcal{I}_{\gamma\gamma} &= \rho^2
        \int_{-T}^T \left|a(t)\right|^2 w(t; \rho) dt, \\
    \label{eq:fisher-autocor-tau-tau}
    \mathcal{I}_{\tau\tau} &= -\rho^2
        \int_{-T}^T \Re\left[a^*(t) \ddot{a}(t)\right] w(t; \rho) dt, \\
    \label{eq:fisher-autocor-gamma-tau}
    \mathcal{I}_{\gamma\tau} &= -\rho^2
        \int_{-T}^T \Im\left[a^*(t) \dot{a}(t)\right] w(t; \rho) dt,
\intertext{where}\label{eq:fisher-weight-factor}
    w(t; \rho) &= \frac{
        \displaystyle
        \exp\left[\frac{\rho^2}{4}\left|a(t)\right|^2\right]
        \left(
        I_0\left[\frac{\rho^2}{4}\left|a(t)\right|^2\right] +
        I_1\left[\frac{\rho^2}{4}\left|a(t)\right|^2\right]
        \right)
    }{
        \displaystyle
        2 \int_{-T}^T
        \exp\left[\frac{\rho^2}{4}\left|a(t')\right|^2\right]
        I_0\left[\frac{\rho^2}{4}\left|a(t')\right|^2\right]
        dt'
    }.
\end{align}
The notation $I_k$ denotes a modified Bessel function of the first kind. Matrix elements that are not listed have values that are implied by the symmetry of the Fisher matrix. Note that the minus signs are correct but a little confusing; despite them, $\mathcal{I}_{\gamma\gamma}, \mathcal{I}_{\tau\tau} \geq 0$ and $\mathcal{I}_{\gamma\tau} \leq 0$. The time integration limits $[-T, T]$ correspond to a flat prior on arrival time or a time coincidence window between detectors.

\begin{figure*}
    \centering
        \includegraphics[width=\columnwidth]{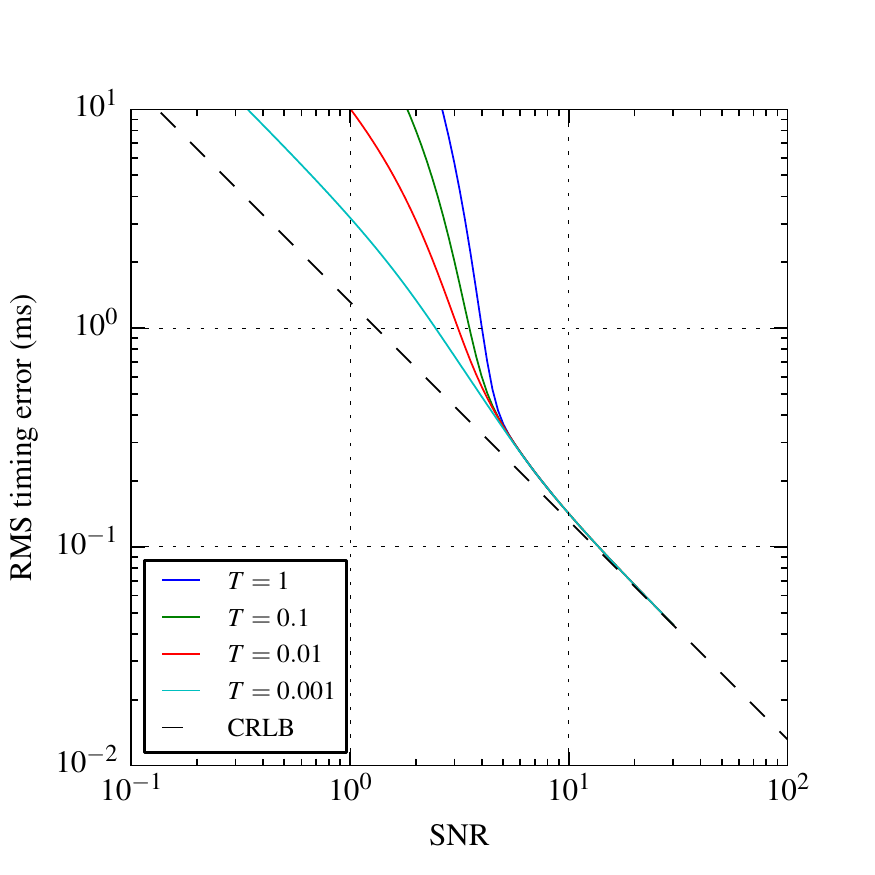}
        \includegraphics[width=\columnwidth]{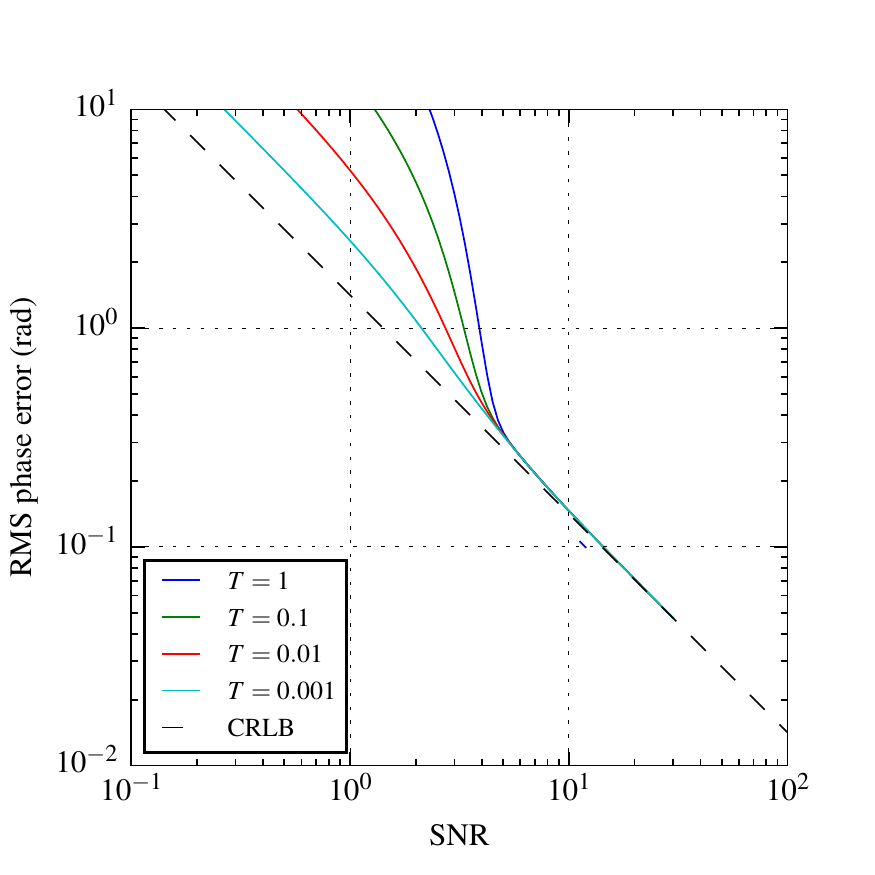}
    \caption[\acl{CRLB} on time and phase accuracy]{\label{fig:crlb-tau}\ac{CRLB} on \acl{RMS} timing uncertainty and phase error, using the likelihood for the full \ac{GW} data [Eq.~(\ref{eq:gaussian-likelihood-spa}); dashed diagonal line] or the autocorrelation likelihood [Eq.~(\ref{eq:autocor-likelihood}); solid lines] with a selection of arrival time priors.}
\end{figure*}

We can show that the weighting function $w(t; \rho)$ approaches a Dirac delta function as $\rho \rightarrow \infty$, so that the Fisher matrix for the autocorrelation likelihood approaches the Fisher matrix for the full \ac{GW} data, Eq.~(\ref{eq:fisher-matrix}), as $\rho \rightarrow \infty$. The Bessel functions asymptotically approach
\begin{equation*}
    I_0(x), I_1(x) \rightarrow \frac{e^x}{\sqrt{2 \pi x}}
    \textrm{ as } x \rightarrow \infty.
\end{equation*}
For large $\rho$, the exponents of $e^{\rho^2}$ dominate Eq.~(\ref{eq:fisher-weight-factor}), and we can write
\begin{equation*}
    w(t; \rho) \rightarrow \frac{
        \displaystyle
        \exp\left[\frac{\rho^2}{2}|a(t)|^2\right]
    }{
        \displaystyle
        \int_{-T}^T \exp\left[\frac{\rho^2}{2}|a(t')|^2\right] dt'
    }
    \textrm{ as } \rho \rightarrow \infty.
\end{equation*}
The Taylor expansion of $|a(t)|^2$ is
\begin{align}
    \nonumber
    |a(t)|^2 &= 1 + \frac{1}{2} \left(\frac{\partial^2}{\partial t^2}|a(t)|^2 \Bigg|_{t=0}\right) t^2 + \mathcal{O}(t^4) \\
    \nonumber
    &= 1 - {\omega_\mathrm{rms}}^2 t^2 + \mathcal{O}(t^4).
\end{align}
Substituting, we find that $w(t; \rho)$ approaches a normalized Gaussian distribution:
\begin{equation*}
    w(t; \rho) \approx \frac{
        \displaystyle
        \exp\left[-\frac{1}{2} \rho^2 {\omega_\mathrm{rms}}^2 t^2\right]
    }{
        \displaystyle
        \int_{-T}^T \exp\left[-\frac{1}{2} \rho^2 {\omega_\mathrm{rms}}^2 (t')^2\right] dt'
    }.
\end{equation*}
And finally, because the Dirac delta function may be defined as the limit of a Gaussian, $w(t; \rho) \rightarrow \delta(t)$ as $\rho \rightarrow \infty$.

\begin{figure}
    \centering
    \includegraphics{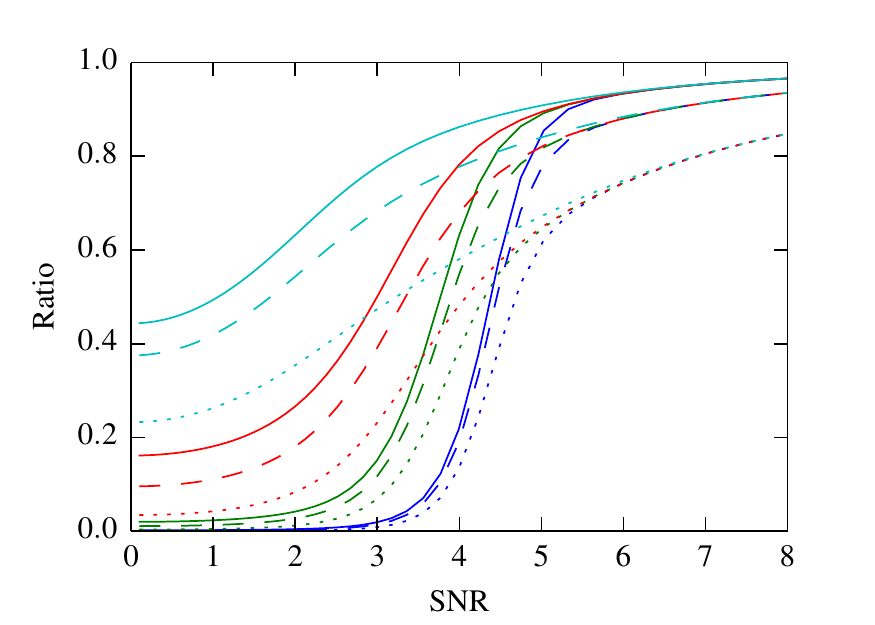}
    \caption[Ratio of Fisher matrix elements between autocorrelation likelihood and full \acs{GW} data]{\label{fig:fishfactor}Ratio between Fisher matrix elements (solid: $\textproto{D}_{\gamma\gamma}$, dashed: $\textproto{D}_{\gamma\tau}$, dotted: $\textproto{D}_{\tau\tau}$) for the autocorrelation likelihood and the full \ac{GW} data. Colors correspond to different arrival time priors as in Fig.~\ref{fig:crlb-tau}.}
\end{figure}

We can now write the Fisher matrix for the autocorrelation likelihood in a way that makes a comparison to the full signal model explicit. Define
\begin{align*}
    \mathcal{I}_{\gamma\gamma} &= \rho^2 \cdot \textproto{D}_{\gamma\gamma}(\rho), \\
    \mathcal{I}_{\tau\tau} &= \rho^2 \overline{\omega^2} \cdot \textproto{D}_{\tau\tau}(\rho), \\
    \mathcal{I}_{\gamma\tau} &= -\rho^2 \overline{\omega} \cdot \textproto{D}_{\gamma\tau}(\rho).
\end{align*}
Now, the $\textproto{D}_{ij}$\footnote{The Fish(er) factor.}
contain the integrals from Eqs.~(\ref{eq:fisher-autocor-gamma-gamma},~\ref{eq:fisher-autocor-tau-tau},~\ref{eq:fisher-autocor-gamma-tau}) and encode the departure of the autocorrelation likelihood from the likelihood of the full data at a low \ac{SNR}. All of the $\textproto{D}_{ij}(\rho)$ are sigmoid-type functions that asymptotically approach 1 as $\rho \rightarrow \infty$ (see Figs.~\ref{fig:crlb-tau} and~\ref{fig:fishfactor}). The transition \ac{SNR} $\rho_\mathrm{crit}$ is largely the same for all three nontrivial matrix elements, and is determined by the time coincidence window $T$ and the signal bandwidth $\omega_\mathrm{rms}$.

In the limit of large \ac{SNR}, our interpretation is that the point estimates $(\hat\rho, \hat\gamma, \hat\tau)$ contain all of the information about the underlying extrinsic parameters.

On the other hand, in the low \ac{SNR} limit, the diminishing value of $\textproto{D}_{ij}(\rho)$ reflects the fact that some information is lost when the full data $\mathbf{x}$ are discarded. Concretely, as the prior interval $T$ becomes large compared to 1/$\rho\omega_\mathrm{rms}$, the \ac{ML} estimator becomes more and more prone to picking up spurious noise fluctuations far from the true signal. Clearly, when the coincidence window $T$ is kept as small as possible, more information is retained in the \ac{ML} point estimates. Put another way, if $T$ is small, then the transition \ac{SNR} $\rho_\mathrm{crit}$ is also small and fainter signals become useful for parameter estimation. In this way, the \ac{BAYESTAR} likelihood exhibits the \emph{threshold effect} that is well known in communication and radar applications \citep{barankin1949locally,mcaulay1969barankin,mcaulay1971barankin}.

In the following sections, we describe our prior and our numerical schemes to integrate over nuisance parameters, which together amount to the \ac{BAYESTAR} algorithm.

\section{Prior and problem setup}
\label{sec:prior}

The detection pipeline supplies a candidate, $\{\left\{ \hat\rho_i, \hat\gamma_i, \hat\tau_i \right\}_i, \hat{\bm\theta}_\mathrm{in}\}$, and discretely sampled noise \acp{PSD}, $S_i(\omega_j)$, for all detectors. We compute the \ac{GW} signal for a source with intrinsic parameters equal to the detection pipeline's estimate, $H(\omega; \hat{\bm\theta}_\mathrm{in})$.
Then, we find the \ac{SNR}=1 horizon distance $r_{1,i}$ for each detector by numerically integrating Eq.~(\ref{eq:horizon}).

We have no explicit prior on the intrinsic parameters; in our analysis they are fixed at their \ac{ML} estimates, $\hat{\bm\theta}_\mathrm{in}$.\footnote{As noted in footnote \ref{footnote:template-banks}, the detection template bank is typically designed to uniformly sample the Jeffreys prior on the intrinsic parameters. Due to the equivalence of marginalization and maximization with respect to a parameter under a Gaussian distribution, fixing the intrinsic parameters at their \ac{ML} estimates is roughly equivalent to selecting the Jeffreys prior.}

The arrival time prior is connected to the origin of the detector coordinate system. Given the Earth\nobreakdashes-fixed coordinates of the detectors $\mathbf{n}_i$ and the arrival times $\tau_i$, we compute their averages weighted by the timing uncertainty formula:
\begin{equation*}
    \langle \mathbf{n} \rangle = \frac{
        \displaystyle
        \sum_i \frac{\mathbf{n}_i}
            {\left(\hat\rho_i \omega_{\mathrm{rms},i}\right)^2}
    }{
        \displaystyle
        \sum_i \frac{1}{\left(\hat\rho_i \omega_{\mathrm{rms},i}\right)^2}
    },
    \qquad
    \langle \hat\tau \rangle = \frac{
        \displaystyle
        \sum_i \frac{\hat\tau_i}
            {\left(\hat\rho_i \omega_{\mathrm{rms},i}\right)^2}
    }{
        \displaystyle
        \sum_i \frac{1}{\left(\hat\rho_i \omega_{\mathrm{rms},i}\right)^2}
    }.
\end{equation*}
Then, we subtract these means:
\begin{equation*}
    \mathbf{n}_i \leftarrow \mathbf{n}_i - \langle \mathbf{n} \rangle,
    \qquad
    \hat\tau_i \leftarrow \hat\tau_i - \langle \hat\tau \rangle.
\end{equation*}
In these coordinates, now relative to the weighted detector array barycenter, the arrival time prior is uniform in $-T \leq t \leq T$, with $T = \max\limits_i |\mathbf{n}_i| / c + 5~\textrm{ms}$.

The distance prior is a user-selected power of distance,
\begin{equation*}
    p(r) \propto \begin{cases}
        r^m & \text{if } r_\mathrm{min} < r < r_\mathrm{max} \\
        0 & \text{otherwise},
    \end{cases}
\end{equation*}
where $m=2$ for a prior that is uniform in volume and $m=-1$ for a prior that is uniform in the logarithm of the distance. If a distance prior is not specified, the default is uniform in volume out to the maximum \ac{SNR}=4 horizon distance:
\begin{equation*}
    m = 2,
    \qquad
    r_\mathrm{min} = 0,
    \qquad
    r_\mathrm{max} = \frac{1}{4} \max_i r_{1,i}.
\end{equation*}

Finally, the prior is uniform in $-1 \leq \cos\iota \leq 1$ and $0 \leq \psi < \pi$.

We compute the autocorrelation function for each detector from $t = 0$ to $t = T$ at intervals of $\Delta t = 1/f_s$, where $f_s$ is the smallest power of 2 that is greater than or equal to the Nyquist rate. Because BNS signals typically terminate at about 1500~Hz, a typical value for $\Delta t$ is $(4096\,\textrm{Hz})^{-1}$. We use a pruned \ac{FFT} because for BNS systems, the \ac{GW} signal remains in LIGO's sensitive band for $\sim$100\nobreakdashes--1000~s, whereas $T \sim 10$~ms.\footnote{See \url{http://www.fftw.org/pruned.html} for a discussion of methods for computing the pruned \ac{FFT}, the first $K$ samples of an \ac{FFT} of length $N$.}

\section{Marginal posterior}
\label{sec:marginal-posterior}

The marginal posterior as a function of the sky location is
\begin{multline}
    f(\alpha, \delta) \propto
    \int_{0}^{\pi}
    \int_{-1}^{1}
    \int_{-T}^{T}
    \int_{r_\mathrm{min}}^{r_\mathrm{max}}
    \int_{0}^{2\pi}
    \\
    \exp \left[ - \frac{1}{2} \sum_i {\rho_i}^2
        + \sum_i \hat\rho_i \rho_i \Re \left\{ e^{i \tilde{\gamma}_i}
        a^*(\tilde{\tau}_i)
        \right\}
    \right] \\
    r^m d\phi_c \, dr \, dt_\oplus \, d\cos{\iota} \, d\psi.
\end{multline}

To marginalize over the coalescence phase, we can write $\tilde{\gamma}_i = \tilde{\gamma}_i^\prime + 2\phi_c$. Then, integrating over $\phi_c$ and suppressing normalization factors, we get
\begin{multline}
    f(\alpha, \delta) \rightarrow
    \int_{0}^{\pi}
    \int_{-1}^{1}
    \int_{-T}^{T}
    \int_{r_\mathrm{min}}^{r_\mathrm{max}}
    \\
    \exp \left[ - \frac{1}{2} \sum_i {\rho_i}^2 \right] I_0 \left[
            \left| \sum_i \hat\rho_i \rho_i e^{i \tilde{\gamma}_i} a_i^*(\tilde{\tau}_i)
            \right|
    \right] \\
    r^m dr \, dt_\oplus \, d\cos{\iota} \, d\psi.
\end{multline}
In the above equation, we need not distinguish between $\tilde{\gamma}_i$ and $\tilde{\gamma}_i^\prime$ because the likelihood is now invariant under arbitrary phase shifts of all of the detectors' signals.

\subsection{Integral over angles and time}

\begin{figure*}
    \begin{minipage}[t]{0.5\textwidth}
        \centering
        (a) Polarization angle \\
        \includegraphics{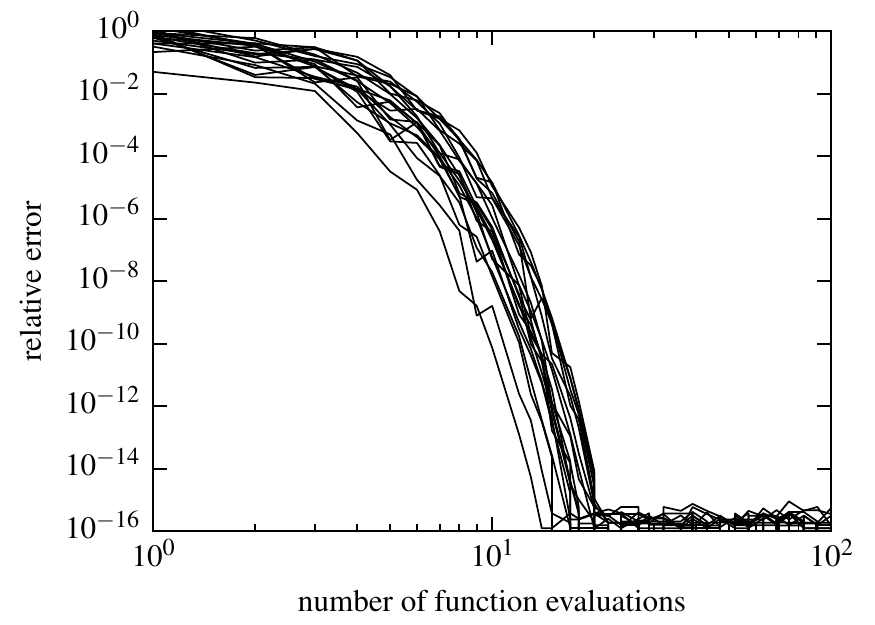}
    \end{minipage}    \begin{minipage}[t]{0.5\textwidth}
        \centering
        (b) Inclination angle \\
        \includegraphics{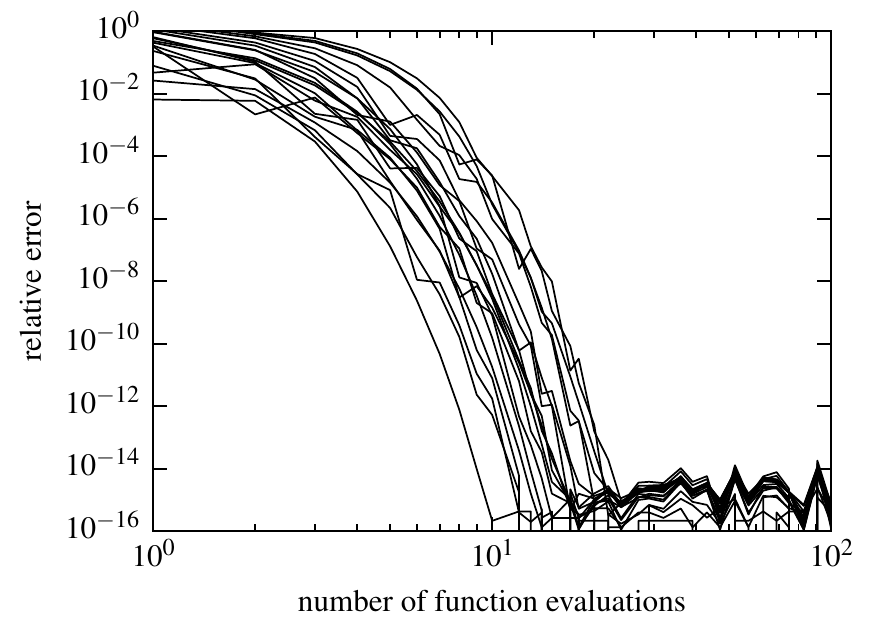}
    \end{minipage}
    \caption{\label{fig:angle-convergence}Relative error in the \ac{BAYESTAR} integration scheme as a function of the number of Gaussian quadrature nodes. The two panels describe (a) the integral over the polarization angle $\psi$ and (b) the integral over the inclination angle $\iota$.}
\end{figure*}

The integrand is periodic in $\psi$, so simple Newton\nobreakdashes--Cotes quadrature over $\psi$ exhibits extremely rapid convergence (see Fig.~\ref{fig:angle-convergence}). We therefore sample the posterior on a regular grid of ten points from 0 to $\pi$.

The integral over $\cos\iota$ converges just as rapidly with Gauss\nobreakdashes--Legendre quadrature (see Fig.~\ref{fig:angle-convergence}), so we use a ten\nobreakdashes-point Gauss\nobreakdashes--Legendre rule for integration over $\cos\iota$.

We sample $t_\oplus$ regularly from $-T$ to $T$ at intervals of $\Delta t$. This is typically $\sim 2 (10\,\mathrm{ms})(4096\,\mathrm{Hz}) \approx 80$~samples. We use Catmull\nobreakdashes--Rom cubic splines to interpolate the real and imaginary parts of the autocorrelation functions between samples.

\subsection{Integral over distance}
\label{sec:distance}

The distance integral is now performed differently from what we initially described in Refs.~\cite{leo-singer-thesis,BerryLocalization}; the method described in the present work is about an order of magnitude faster. We define $\rho_i = \omega_i / r$ in order to absorb all of the distance-independent terms in the amplitudes into $\omega_i$ and then define
\begin{align}
    p^2 &= \frac{1}{2} \sum_i {\omega_i}^2 \label{eq:likelihood-p-factor} \\
    b &= \left| \sum_i \hat\rho_i \omega_i e^{i \tilde{\gamma}_i}
        a_i^*(\tilde{\tau}_i) \right|. \label{eq:likelihood-b-factor} \end{align}
The innermost integral over distance $r$ may then be written as
\begin{align}\label{eq:distance-integral}
    \mathscr{F} &=
        \int_{\mathrlap{r_\mathrm{min}}}^{\mathrlap{r_\mathrm{max}}}
        \exp\left[-\frac{p^2}{r^2}\right]
        I_0\left[\frac{b}{r}\right] r^m dr \nonumber \\
    &=
        \int_{\mathrlap{r_\mathrm{min}}}^{\mathrlap{r_\mathrm{max}}}
        \exp\left[-\frac{p^2}{r^2}+\frac{b}{r}\right]
        \overline{I}_0\left[\frac{b}{r}\right] r^m dr
\end{align}
or, completing the square,
\begin{align}
    \mathscr{F} &= \exp\left[\frac{p^2}{{r_0}^2}\right]
        \int_{\mathrlap{r_\mathrm{min}}}^{\mathrlap{r_\mathrm{max}}}
        \exp\left[-\left(\frac{p}{r} - \frac{p}{r_0}\right)^2\right]
        \overline{I}_0\left[\frac{2p^2}{r r_0}\right] r^m dr \\
    &= \exp\left[\frac{p^2}{{r_0}^2}\right] \mathscr{G},
\end{align}
where
\begin{align}
    r_0 &= 2 p^2 / b \\
    \overline{I}_0(x) &= \exp(-|x|) I_0(x).
\end{align}
The coefficients $p^2$ and $b$ are non\nobreakdashes-negative and independent of distance. $p$ has a maximum value of
\begin{equation}
    p_\mathrm{max} = \sqrt{\frac{1}{2} \sum_i \left(\frac{r_{1,i}}{r_\mathrm{max}}\right)^2}.
\end{equation}
The symbol $\overline{I}_0$ denotes an exponentially scaled Bessel function. In the limit of large argument, $I_0(|x|) \sim \exp(|x|) / \sqrt{2 \pi |x|}$ \citep{Olver:2010:NHMF,NIST:DLMF}\footnote{\url{http://dlmf.nist.gov/10.40.E1}}. The scaled Bessel function is useful for evaluation on a computer because it has a relatively small range $(0, 1]$ and varies slowly in proportion to $x^{1/2}$.

\subsubsection{Parameter grid}

\begin{figure}
    \includegraphics{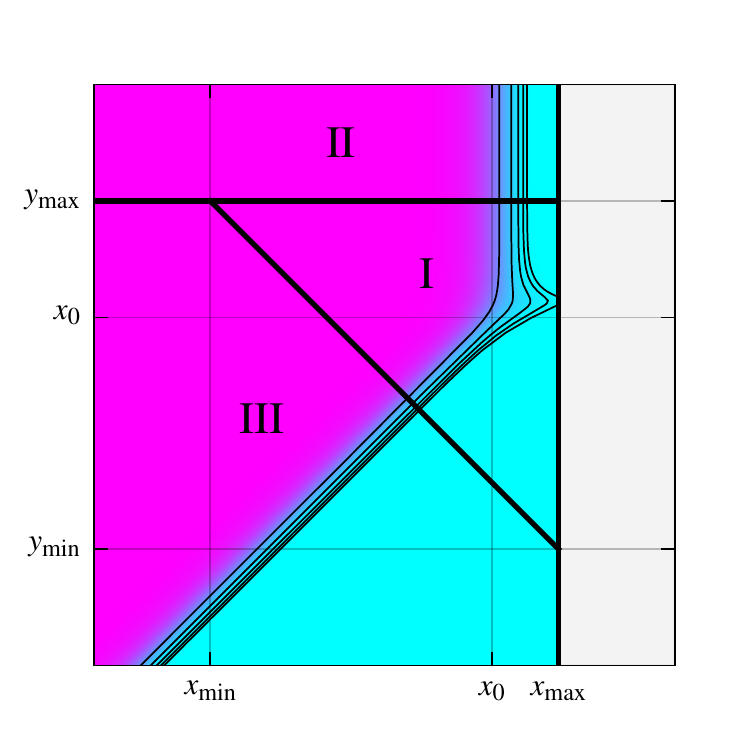}
    \caption{\label{fig:radial-integral-interpolant}Partition of the parameter space of the distance integral into three regions for (bi)cubic interpolation.}
\end{figure}

This integral is not particularly amenable to low-order Gaussian quadrature. However, luckily $\mathscr{G}$ is a very well-behaved function of $p$ and $r_0$, so we evaluate it using a lookup table and bicubic interpolation. The lookup table is produced in logarithmic coordinates
\begin{equation}
    x = \log p, \qquad y = \log r_0.
\end{equation}
As shown in Fig.~\ref{fig:radial-integral-interpolant}, the function basically consists of a plateau region in the upper-left half of the plane delimited by the lines $y = x$ and $x = \log p_0$, with
\begin{equation}
    p_0 = \frac{1}{2}
    \begin{cases}
        r_\mathrm{max} & \text{if } m \geq 0 \\
        r_\mathrm{min} & \text{if } m < 0.
    \end{cases}
\end{equation}
We tabulate $\mathscr{G}$ on a $400 \times 400$ regular grid spanning the range
\begin{align}
    x_0 &= \log \min(p_0, p_\mathrm{max}) \\
    x_\mathrm{min} &= x_0 - (1 + \sqrt{2}) \alpha \\
    x_\mathrm{max} &= \log p_\mathrm{max} \\
    y_\mathrm{min} &= 2 x_0 - \sqrt{2} \alpha - x_\mathrm{max} \\
    y_\mathrm{max} &= x_0 + \alpha
\end{align}
where $\alpha = 4$ is a constant parameter that determines the extent of the grid.

\subsubsection{Lookup table construction}

The lookup table for $\mathscr{G}$ is populated as follows. If we neglect both the Bessel function and the $r^m$ prior, then the approximate likelihood $\exp(-(p/r-p/r_0)^2)$ is maximized when $r = r_0$. The likelihood takes on a factor $\eta$ (say, $\eta=0.01$) of its maximum value when
\begin{equation}
    r = r_\pm = \left(\frac{1}{r_0} \mp \frac{\sqrt{-\log\eta}}{p}\right)^{-1}.
\end{equation}

\begin{figure}
    \begin{center}
        \includegraphics[width=\columnwidth]{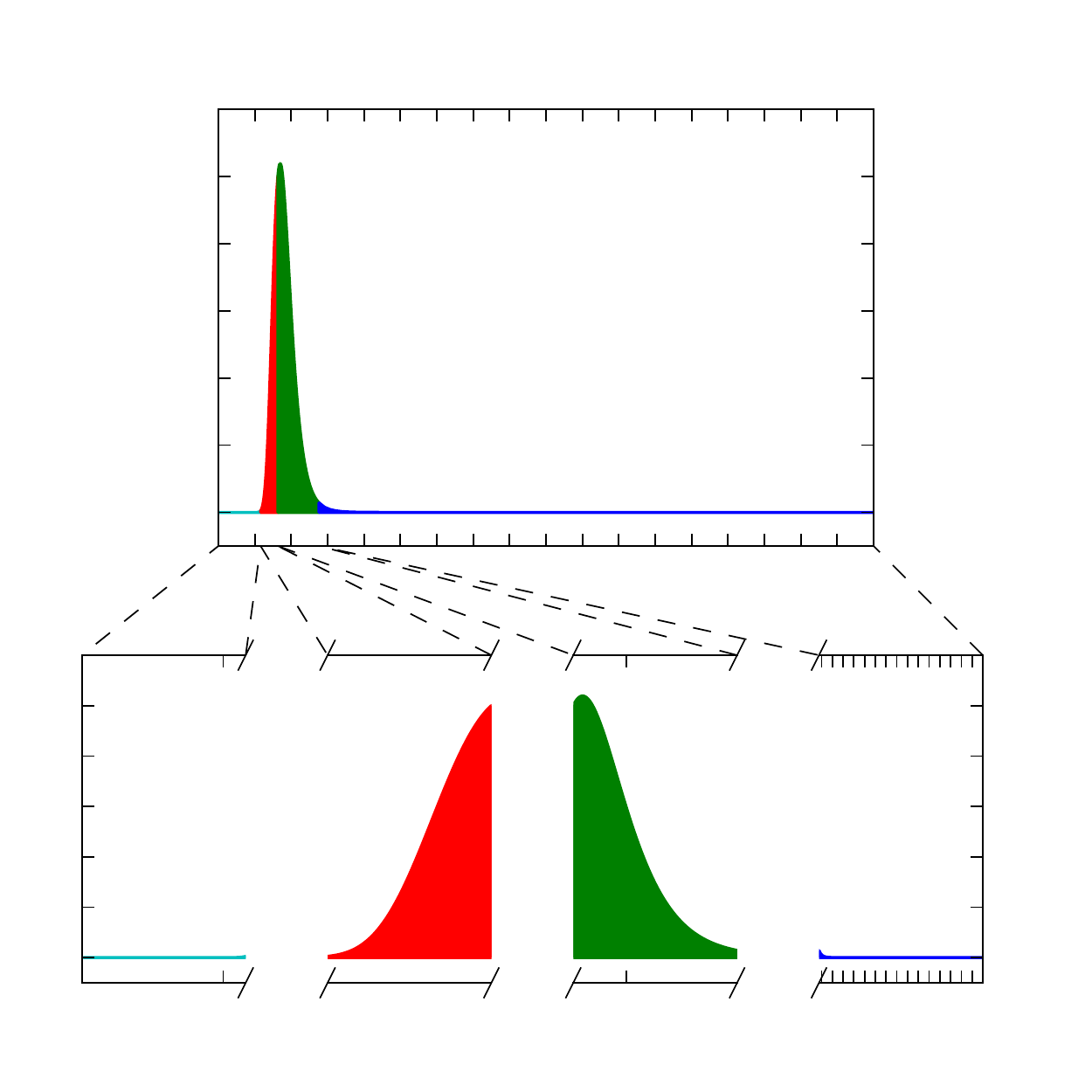}
    \end{center}
    \caption{\label{fig:radial_integrand}Illustration of initial subdivisions for the distance integration scheme. The distance increases from left to right. In the color version, the left\nobreakdashes-hand tail, the left\nobreakdashes- and right\nobreakdashes-hand sides of the maximum likelihood peak, and the right\nobreakdashes-hand tail, are colored cyan, red, green, and blue, respectively.}
\end{figure}

We have now identified up to five breakpoints that partition the distance integrand into up to four intervals with quantitatively distinct behavior. These intervals are depicted in Fig.~\ref{fig:radial_integrand} with the distance increasing from left to right. There is a left\nobreakdashes-hand or small distance tail in which the integrand is small and monotonically increasing, a left\nobreakdashes- and right\nobreakdashes-hand side of the maximum likelihood peak, and a right\nobreakdashes-hand tail in which the integrand is small and monotonically decreasing. These breakpoints are
\begin{equation}
    r_\mathrm{break} = \{ r \in
    \left\{
    \begin{array}{c}
    r_\mathrm{min} \\
    r_- \\
    r_0 \\
    r_+ \\
    r_\mathrm{max}
    \end{array}
    \right\} :
    r_\mathrm{min} \leq r
    \leq r_\mathrm{max}\}.
\end{equation}

We use these breakpoints as initial subdivisions in an adaptive Gaussian quadrature algorithm\footnote{For instance, \acl{GSL}'s \texttt{gsl\_integrate\_qagp} function, \url{http://www.gnu.org/software/gsl/manual/html_node/QAGP-adaptive-integration-with-known-singular-points.html}.}. This function estimates the integral over each subdivision and each interval's contribution to the total error, then subdivides the interval that contributes the most to the error. Subdivisions continue until a fixed total fractional error is reached. In this way, most integrand evaluations are expended on the most important distance interval, whether that happens to be the tails (when the posterior is dominated by the prior) or the peak (when the posterior is dominated by the observations).

\subsubsection{Interpolation}

\begin{figure}
    \begin{center}
        \includegraphics[width=\columnwidth]{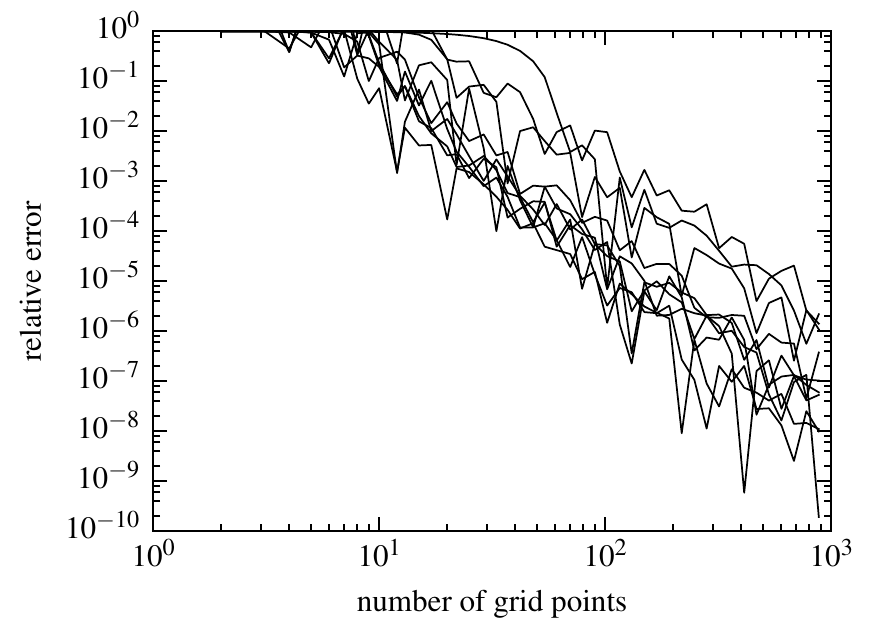}
    \end{center}
    \caption{\label{fig:distance_interpolant_convergence}Relative error in the \ac{BAYESTAR} distance integral interpolation scheme as a function of the size of the grid.}
\end{figure}

The interpolant is evaluated slightly differently depending on which of the three regions marked I, II, and III in Fig.~\ref{fig:radial-integral-interpolant} contains the point of interest. In region I, we use bicubic interpolation of $\log \mathcal{G}$ in $x$ and $y$. In region II, we use univariate cubic interpolation of $\log \mathcal{G}$ in $x$, with the sample points taken from the horizontal boundary between regions I and II. In region III, we use univariate cubic interpolation of $\log \mathcal{G}$ in $u = (x - y) / 2$, with the sample points taken from the downward diagonal boundary between regions I and III. Finally, the distance integral $\mathscr{F}$ is obtained by multiplying the interpolated value of $\mathscr{G}$ by $\exp\left(p^2/{r_0}^2\right)$. For a $400\times400$ grid, the entire lookup table scheme is accurate to a relative error of about $10^{-5}$ in $\mathscr{F}$ (see Fig.~\ref{fig:distance_interpolant_convergence}).

\section{Adaptive HEALPix sampling}
\label{sec:adaptive-sampling}

We have explained how we evaluate the marginal posterior at a given sky location. Now we must specify where we choose to evaluate it.

Our sampling of the sky relies completely on \ac{HEALPix}~\cite{healpix}, a special data structure designed for all\nobreakdashes-sky maps. \ac{HEALPix} divides the sky into equal\nobreakdashes-area pixels. There is a hierarchy of \ac{HEALPix} resolutions. A \ac{HEALPix} resolution may be designated by its order $N$. The $N=0$th order or base tiling has 12 pixels. At every successive order, each tile is subdivided into four new tiles. A resolution may also be referred to by the number of subdivisions along each side of the base tiles, $N_\mathrm{side} = 2^N$. There are $N_\mathrm{pix} = 12 {N_\mathrm{side}}^2$ pixels at any given resolution. The \ac{HEALPix} projection uniquely specifies the coordinates of the center of each pixel by providing a mapping from the resolution and pixel index $(N_\mathrm{side}, i_\mathrm{pix})$ to right ascension and declination $(\alpha, \delta)$.

\begin{figure*}
    \includegraphics[width=\textwidth]{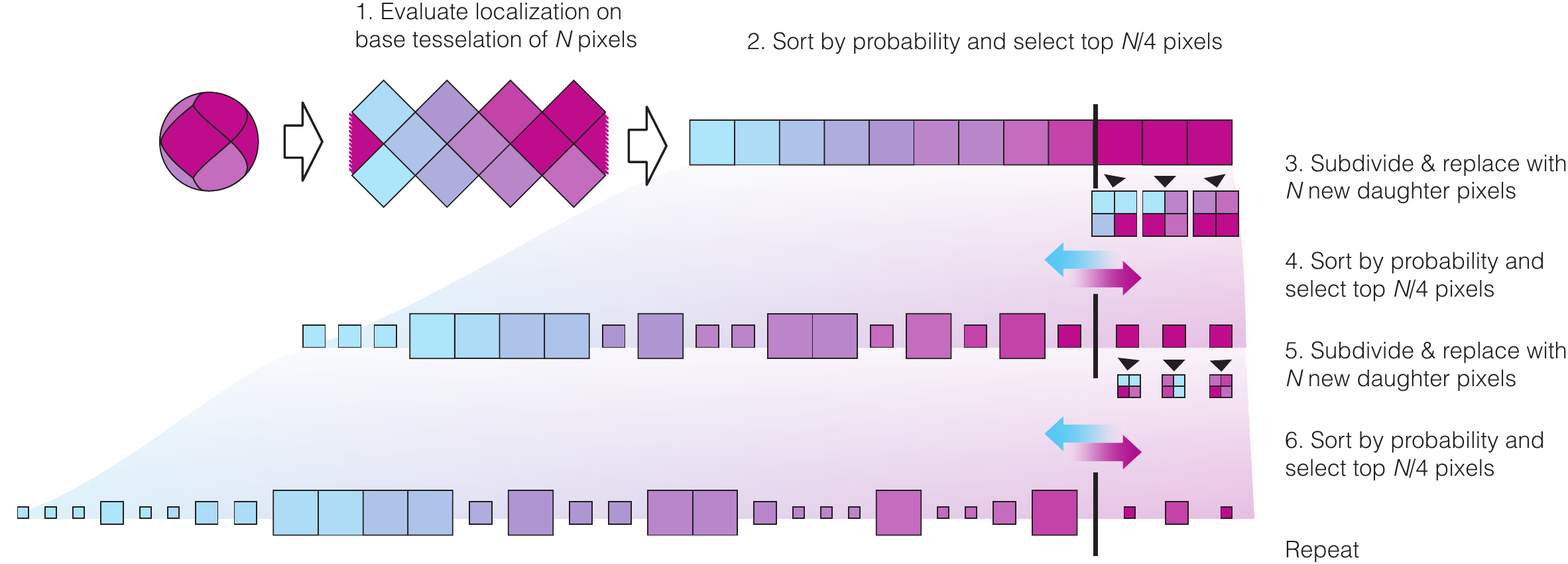}
    \caption{\label{fig:adaptive-illustration}Illustration of the \ac{BAYESTAR} adaptive HEALPix sampling scheme.}
\end{figure*}

The \ac{BAYESTAR} adaptive sampling process works as follows. We begin by evaluating the posterior probability density at the center of each of the $N_{\mathrm{pix},0} = 3072$ pixels of an $N_{\mathrm{side},0}=16$ \ac{HEALPix} grid. At this resolution, each pixel has an area of 13.4~deg$^2$. We then rank the pixels by contained probability (assuming constant probability density within a pixel) and subdivide the most probable $N_{\mathrm{pix},0}/4$ pixels into $N_{\mathrm{pix},0}$ new daughter pixels. We then evaluate the posterior again at the centers of the new daughter pixels, sort again, and repeat seven times. By the end of the last iteration, we have evaluated the posterior probability density a total of $8 N_{\mathrm{pix},0}$ times. On \emph{most} subdivision steps, we descend one level deeper in \ac{HEALPix} resolution. This process is illustrated in Fig.~\ref{fig:adaptive-illustration}.

\begin{figure}
    \includegraphics[width=\columnwidth]{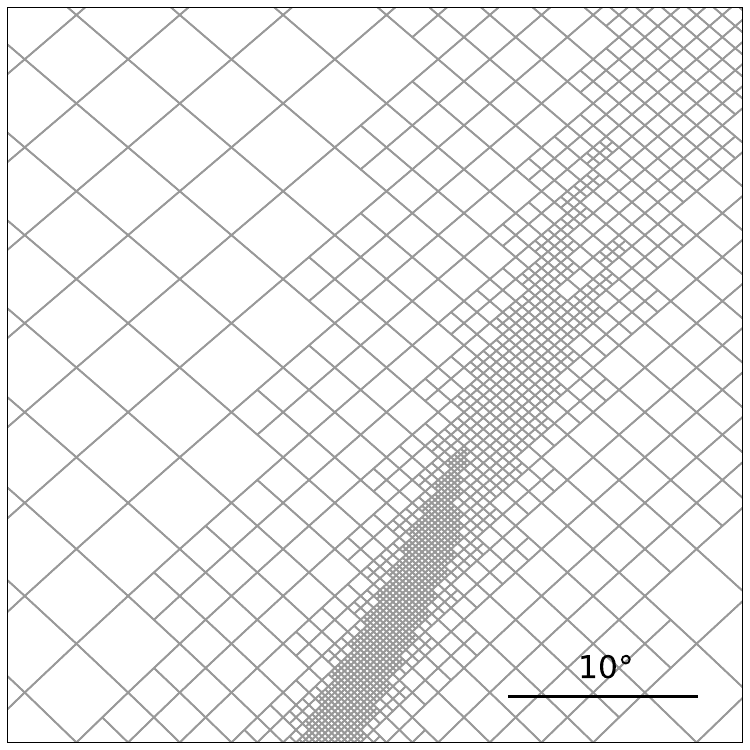}
    \caption{\label{fig:adaptive-mesh}An example multiresolution \ac{HEALPix} mesh arising from the \ac{BAYESTAR} sampling scheme (plotted in a cylindrical projection). This is event 18951 from Ref.~\cite{FirstTwoYears}.}
\end{figure}

The resulting map is a tree structure that describes a mesh of pixels with different resolutions. An example \ac{BAYESTAR} subdivision is shown in Fig.~\ref{fig:adaptive-mesh}. To convert this mesh into a \ac{FITS}~\cite{FITS} image, we traverse the tree and flatten it into the highest resolution represented. The highest possible resolution is $N_\mathrm{side}=2^{11}$, with an area of $\approx 10^{-3}$~deg$^2$ per pixel.\footnote{Although the resulting sky map contains $N_\mathrm{pix} \approx 5\times10^6$ pixels, at most $\approx 2\times10^4$ pixels have distinct values. For the purpose of delivery to observers, therefore, the output is always \texttt{gzip}\nobreakdashes-compressed with a ratio of $\approx 250:1$.}

\section{Parallelization}

\ac{MCMC} and similar stochastic schemes are typically very resistant to parallelization. However, \ac{BAYESTAR} is completely deterministic and easily parallelizable because each pixel can be evaluated independently from all of the others. \ac{BAYESTAR} consists of nine computationally intensive loops: the generation of the distance integral lookup table and the eight loops over pixels in the adaptive HEALPix sampling step. The iterations of each loop are distributed across multiple cores using OpenMP\footnote{\url{http://openmp.org/}.} compiler directives. In Sec.~\ref{sec:run-time}, we will show that \ac{BAYESTAR}'s run time is almost perfectly proportional to the number of cores, demonstrating that the serial sections (the sorts between the adaptation steps) are a negligible contribution to the overall wall clock time.

\section{Case study}
\label{sec:case-study}

We have completed our description of the \ac{BAYESTAR} algorithm. In Ref.~\cite{FirstTwoYears}, the authors presented a comprehensive and astrophysically realistic sample of simulated \ac{BNS} mergers. We focused on the first two planned Advanced LIGO and Virgo observing runs as desribed in Ref.~\cite{LIGOObservingScenarios}. That work presented a catalog of 500 sky localizations from \ac{BAYESTAR} and LALINFERENCE and dealt with the quantitative position reconstruction accuracy as well as the qualitative sky morphologies. In the present work, we will use the same data set but instead focus on demonstrating the correctness and performance of the \ac{BAYESTAR} algorithm.

\subsection{Observing scenarios}

To review the assumptions made in Ref.~\cite{FirstTwoYears}, the two scenarios are:

\textit{2015}.---The first Advanced \acs{LIGO} observing run, or ``O1,'' scheduled to start in September 2015 and continue for three months. There are only two detectors participating in this run: \acs{LIGO} Hanford (H) and \acs{LIGO} Livingston (L). Both detectors are expected to operate with a direction\nobreakdashes-averaged \ac{BNS} merger range of 40\nobreakdashes--80\,Mpc (though ongoing Advanced \acs{LIGO} commissioning suggests that the higher end of this range will be achieved). As a result of having only two detectors, most localizations are long, thin arcs a few degrees wide and tens to hundreds of degrees long. The median 90\% credible area is about 600\,deg$^2$.

\textit{2016}.---The second observing run, ``O2,'' with the two between Advanced \acs{LIGO} detectors, upgraded to a \ac{BNS} range of 80\nobreakdashes--120\,Mpc, operated jointly with the newly commissioned Advanced Virgo detector (V), operating at a range of 20\nobreakdashes--60\,Mpc. The run is envisioned as lasting for six months in 2016\nobreakdashes--2017. The detectors are assumed to have random and independent 80\% duty cycles. Consequently, all three detectors (HLV) are in science mode about half of the time, with the remaining time divided roughly equally between each of the possible pairs (HL, HV, or LV) and one or fewer detectors (at least two \ac{GW} facilities are required for a detection). Virgo's range is assumed to be somewhat less than \acs{LIGO}'s because its commissioning time table is about a year behind. Although the simulated signals are generally too weak in Virgo to trigger the matched\nobreakdashes-filter pipeline and contribute to \emph{detection}, even these subthreshold signals aid in position reconstruction with LALINFERENCE by lifting degeneracies. As a result, the median 90\% credible area decreases to about 200\,deg$^2$.

All simulated sources have component masses distributed uniformly between 1.2 and 1.6\,$M_\odot$ and randomly oriented spins with dimensionless magnitudes $\chi = c|\mathbf{S}|/Gm^2$ between -0.05 and +0.05. Sky positions and binary orientations are random and isotropic. Distances are drawn uniformly from ${D_\mathrm{L}}^3$, reflecting a uniform source population (neglecting cosmological effects, which are small within the Advanced \acs{LIGO} \ac{BNS} range).

\subsection{Detection and localization}

The simulated waveforms were deposited in Gaussian noise that has been filtered to have the \acp{PSD} consistent with Ref.~\cite{LIGOObservingScenarios}. They were detected using the real\nobreakdashes-time matched\nobreakdashes-filter pipeline, GSTLAL\_INSPIRAL \citep{Cannon:2011vi}. Candidates with estimated \acp{FAR} less than $10^{-2}$\,yr$^{-1}$ were considered to be ``detections.'' Because using Gaussian noise results in lower \acp{FAR} than would be calculated in realistically glitchy detector noise, we imposed an additional detection threshold on the network \ac{SNR}, $\hat{\rho} \geq 12$, which has been found to correspond to a comparable \ac{FAR} in the initial \ac{LIGO} runs.\footnote{See Ref.~\cite{BerryLocalization} for an analysis of the effect of glitchy noise on detection and parameter estimation.} Localizations for the detections were generated with \ac{BAYESTAR} as well as the functionally equivalent and interchangeable LALINFERENCE\_MCMC, LALINFERENCE\_NEST, and LALINFERENCE\_BAMBI samplers (collectively referred to as LALINFERENCE).

\subsection{Areas}

We measured sky localization areas for each event as follows. First, we ranked the \ac{HEALPix} pixels by descending posterior probability. Then, we computed the cumulative sum of the pixels in that order. Finally, we searched for the index of the pixel of which the cumulative sum was equal to a given value: for example, 0.9 if we are interested in the 90\% credible area. That pixel index times the area per pixel is the area of the smallest region of the specified credible level. This area can be thought of as measuring the precision of the sky localization: it is a measure of the scale of the posterior distribution.

We can construct a second measure, called the searched area, as the smallest such constructed area that contains the true location of the source. A telescope with a \ac{FOV} that is small compared to the characteristic scale of the posterior would intercept the true location of the source after covering the searched area. This measure is mainly useful because it measures the accuracy of the sky localization independently of the precision. In other words, it treats the sky map as merely a ranking statistic.

\begin{figure*}
    \includegraphics{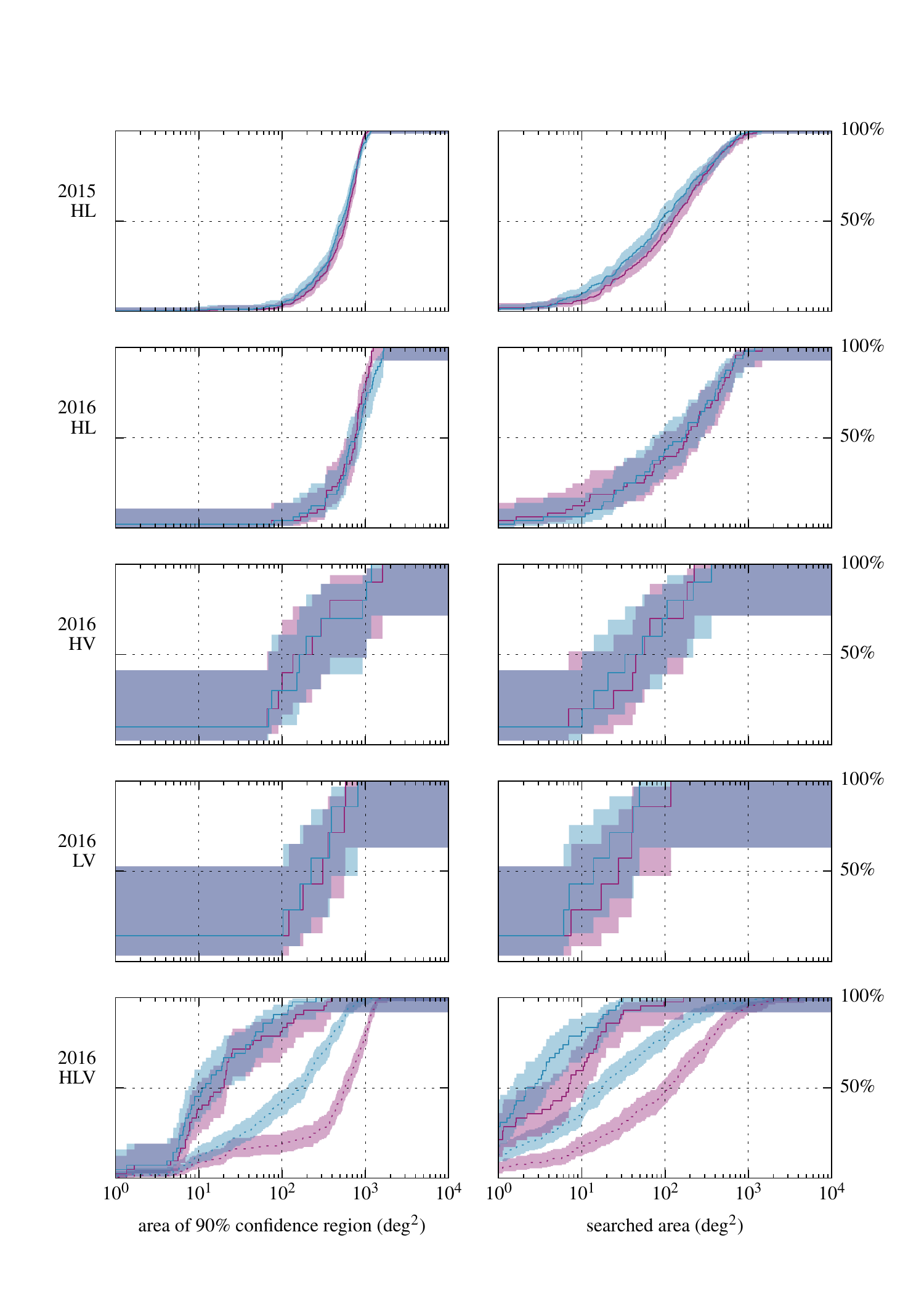}
    \caption{\label{fig:area-hist}Cumulative histograms of sky area, broken down by observing run and detector network. The plots in the left column show the 90\% credible area and the plots in the right column show the searched area. From top bottom, the rows refer to the following observing scenarios/network configurations: 2015/HL, 2016/HL, 2016/HV, 2016/LV, and 2016/HLV. The shaded regions represent the 95\% confidence bounds. The magenta lines represent \ac{BAYESTAR} and the blue lines LALINFERENCE. Where relevant, dotted lines show all events in the given network configuration and solid lines show only events for which the matched\nobreakdashes-filter pipeline triggered on all operating detectors. Note that statistically significant differences in areas between the \ac{BAYESTAR} and LALINFERENCE localizations occur only for events that were below the detection threshold in one or more detectors.}
\end{figure*}

Histograms of the 90\% credible area and the searched area are shown in Fig.~\ref{fig:area-hist}, broken down by observing scenario (2015 or 2016) and detector network (HL, HV, LV, or HLV). Note that there are no statistically significant differences in areas between \ac{BAYESTAR} and LALINFERENCE, with the exception in the 2016/HLV configuration, for which some LALINFERENCE sky maps span about an order of magnitude less area than \ac{BAYESTAR}. If we consider only events for which all three detectors contained a signal that was loud enough to trigger the matched\nobreakdashes-filter pipeline, the difference becomes much smaller and insignificant within 95\% error bars. This is because if the signal is too weak to trigger the detection pipeline in one of the detectors, then \ac{BAYESTAR} receives no information about that detector. This issue does not occur in the two\nobreakdashes-detector configurations (HL, HV, or LV) because two or more triggers are required to report a detection candidate.

This is a significant issue for the 2016 configuration, because the most accurate localizations are possible when all three detectors are operating. However, there may be a simple remedy. As we noted in Sec.~\ref{sec:properties}, the \ac{BAYESTAR} likelihood can be modified to use, instead of the times, phases, and amplitudes on arrival, the full complex matched\nobreakdashes-filter time series from all detectors. The detection pipeline, GSTLAL\_INSPIRAL, would have to be modified to save and transmit a small interval of the complex \ac{SNR} time series (perhaps a few tens of milliseconds) around the time of each detection candidate. In addition to supplying the missing information for subthreshold signals, this would make \ac{BAYESTAR} mathematically equivalent to the LALINFERENCE analysis, but with the intrinsic parameters fixed to their maximum\nobreakdashes-likelihood values. This idea will be pursued in future work.

\subsection{Self-consistency}

As we observed above, the area of a given credible region describes the precision of the sky localization, whereas the searched area describes the accuracy. However, self-consistency requires that the two are related. For example, we should find that on average 90\% of events have their true locations contained within their respective 90\% credible regions. More generally, if we make a cumulative histogram of the credible levels corresponding to the searched areas of all of the events, then we should obtain a diagonal line (with small deviations due to finite sample size). This test, popularized for \ac{GW} data analysis by Ref.~\cite{SiderySkyLocalizationComparison}, is a necessary but not sufficient condition for the validity of any Bayesian parameter estimation scheme.

\begin{figure*}
    \begin{center}
        \includegraphics[width=0.45\textwidth]{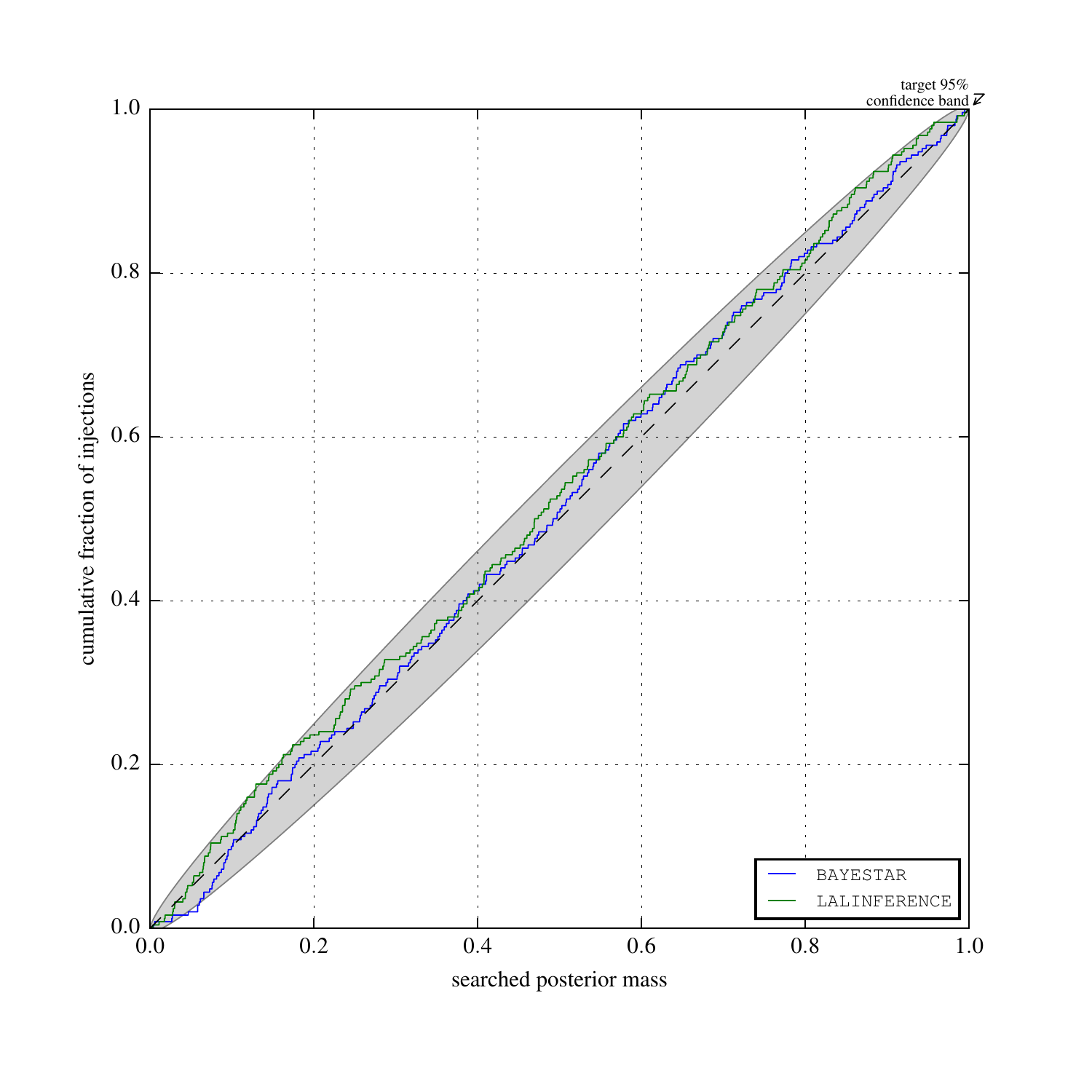}
        \includegraphics[width=0.45\textwidth]{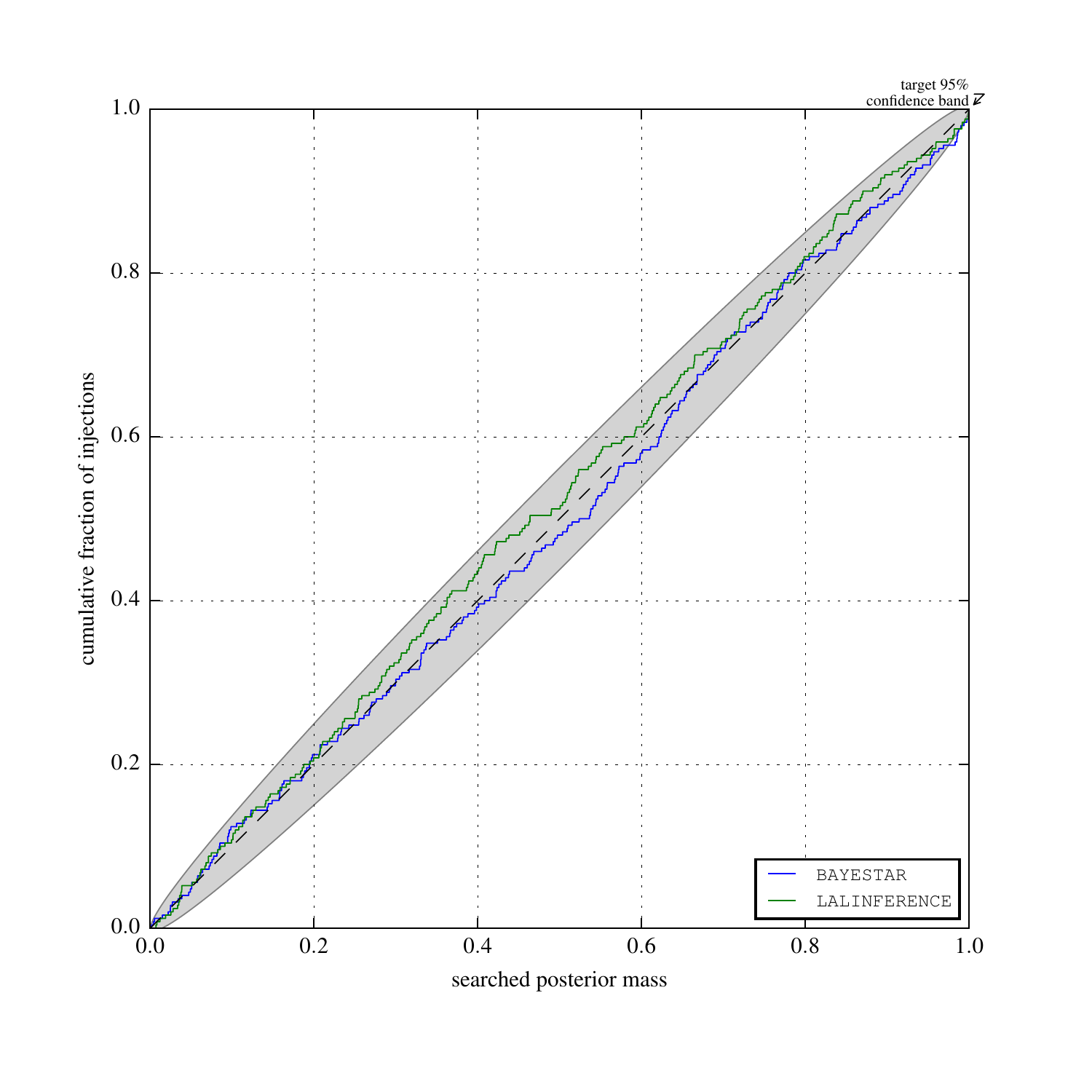}
    \end{center}
    \caption[\acs{BAYESTAR} $P$\nobreakdashes--$P$ plots]{\label{fig:pp}$P$\nobreakdashes--$P$ plots for \ac{BAYESTAR} and LALINFERENCE localizations in the 2015 and 2016 configurations. The gray lozenge around the diagonal is a target 95\% confidence band derived from a binomial distribution.}
\end{figure*}

It is already well established that LALINFERENCE localizations satisfy the $P$\nobreakdashes--$P$ plot test when deployed with accurate templates and reasonable priors. We found at first that \ac{BAYESTAR}'s $P$\nobreakdashes--$P$ plots tended to sag below the diagonal, indicating that though the accuracy (i.e., searched area) was comparable to LALINFERENCE, the \emph{precision} was overstated, with confidence intervals that were only about 70\% of the correct area. This was rectified by prescaling the \acp{SNR} from GSTLAL\_INSPIRAL by a factor of 0.83 prior to running \ac{BAYESTAR}. This correction factor suggests that, for example, a \ac{SNR}=10 trigger from GSTLAL\_INSPIRAL has the \emph{effective information content} of a \ac{SNR}=8.3 signal. The missing information may be due to losses from the discreteness of the template bank, from the \acl{SVD}, from mismatch between the matched\nobreakdashes-filter templates and the simulated signals, from the small but nonzero correlations between masses and intrinsic parameters, or from elsewhere within the detection pipeline. The correction is hard coded into the rapid localization. With it, the $P$\nobreakdashes--$P$ plots are diagonalized without negatively affecting the searched area (see Fig.~\ref{fig:pp}).

\subsection{Run time}
\label{sec:run-time}

\begin{figure}
    \centering
    \includegraphics{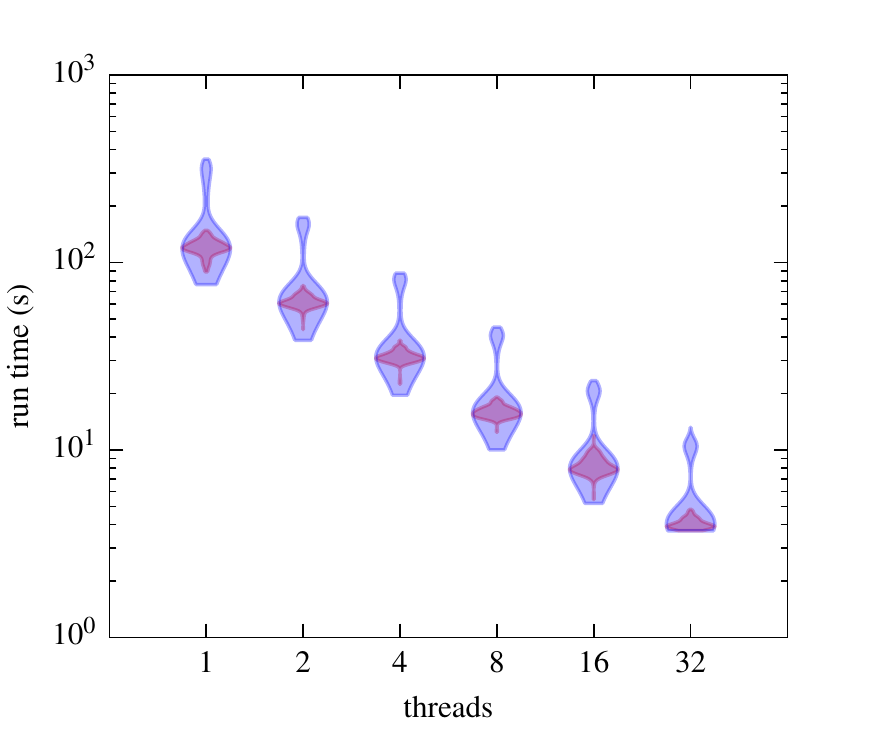}
    \caption[\acs{BAYESTAR} run time]{\label{fig:runtimes}Violin plot of \ac{BAYESTAR} run times as the number of OpenMP threads is varied from 1 to 32. The 2015 scenario is shown in red and the 2016 scenario in blue.}
\end{figure}

Since \ac{BAYESTAR} is designed as one of the final steps in the real\nobreakdashes-time \ac{BNS} search, it is important to characterize how long it takes to calculate a sky map. We compiled \ac{BAYESTAR} with the Intel~C~Compiler~(\texttt{icc}) at the highest architecture\nobreakdashes-specific optimization setting (\texttt{-ipo~-O3~-xhost}). We timed it under Scientific~Linux~6.1 on a Supermicro SuperServer 6028TP\nobreakdashes-HTTR system with dual 8\nobreakdashes-core Intel Xeon E5\nobreakdashes-2630~v3 CPUs clocked at 2.40\,GHz, capable of executing 32 threads simultaneously (with hyperthreading). In Fig.~\ref{fig:runtimes}, we show how long it took to calculate a localization with \ac{BAYESTAR} as the number of OpenMP threads was varied from 1 to 32. This is a violin plot, a smoothed vertical histogram. The magenta regions show run times for a two\nobreakdashes-detector network (HL) modeled on the first scheduled Advanced \ac{LIGO} observing run in 2015, and the blue regions show run times for a three\nobreakdashes-detector network (HLV) based on the second planned observing run in 2016. These are the two observing scenarios that are discussed in Ref.~\cite{FirstTwoYears}.

Several features are apparent. First, at any number of threads, the two configurations have similar run times, although the 2016 events contain a subpopulation of outliers that take about 2.5 times as long as the 2015 events. These are probably due to taking one of the more expensive code branches in the distance integral interpolation. Second, the run times decrease proportionally to the number of threads. Based on experiences running \ac{BAYESTAR} on the 32\nobreakdashes-core (64 threads with hyperthreading) cluster login machine, we expect the almost ideal parallel speedup to continue on machines with even more processors.

With just one thread, the \ac{BAYESTAR} analysis takes 76\nobreakdashes--356~s, already orders of magnitude faster than the full parameter estimation. With 32 threads, \ac{BAYESTAR} takes just 4\nobreakdashes--13~s. In practice, \ac{BAYESTAR}'s data handling (reading the detectors' \acp{PSD}, communicating with the \ac{GW} candidate database, writing \ac{FITS} files) takes an additional 15~s or so, though this overhead could be reduced by parallelizing many of these steps. The overall latency is comparable to the other stages (data aggregation, trigger generation, alert distribution) in the real\nobreakdashes-time \ac{BNS} analysis; therefore, any significant further speedup would require significant changes through Advanced \acs{LIGO} computing and infrastructure. The 32\nobreakdashes-thread configuration is representative of how \ac{BAYESTAR} might be deployed in early Advanced~\ac{LIGO}.\footnote{\ac{BAYESTAR} has been successfully ported to the Intel's Many Integrated Core architecture and has been tested in a 500 thread configuration on a system with dual Intel Xeon Phi 5110P coprocessors.} For comparison, sky localization with LALINFERENCE takes about 100\,h~\cite{BerryLocalization}.

Note that this benchmark shows \ac{BAYESTAR} to be an order of magnitude faster than what was reported in Refs.~\cite{leo-singer-thesis,BerryLocalization} due to the changes in the distance integration scheme that we noted in Sec.~\ref{sec:distance}.

\section{Future work}

One immediately pressing direction for future work is to address the issue of subthreshold signals, as this will be a major issue when Advanced Virgo comes online in 2016\nobreakdashes--2017. Using the full \ac{SNR} time series in place of the autocorrelation function seems like a promising avenue; implementing this requires some infrastructure changes to both the matched-filter pipeline and \ac{BAYESTAR}. Along these lines, we also refer the reader to Ref.~\cite{PankowRapidPE} for a similar, non\nobreakdashes-\ac{MCMC} approach to the rapid exploration of the full parameter space.

A more open\nobreakdashes-ended question is how to account for spin precession. The simulations in Ref.~\cite{FirstTwoYears} and in this paper featured extremely modest spins of $\chi \leq 0.05$, consistent with the fastest known pulsars in binaries \citep{2003Natur.426..531B,DetectingBNSSystemsWithSpin}. The signals were detected using a template bank that lacked spins entirely. Reference~\cite{FirstTwoYearsSpin} shows that using nonspinning \ac{BNS} templates for parameter estimation has negligible impact on sky localization. However, if one or both companions are spinning as fast as a millisecond pulsar, $\chi \sim 0.4$ \cite{FastestSpinningMillisecondPulsar}, or even near breakup, $\chi \sim 0.7$, then the orbital plane may precess; in this case, spins can no longer be neglected for detection \cite{DetectingBNSSystemsWithSpin} and may also be important for parameter estimation. Since spinning \ac{BNS} searches are still an active area of development, \ac{BAYESTAR}'s sky localization accuracy in this regime should be reexamined in the future.

Although the response time of \ac{BAYESTAR} has been driven by the anticipated time scales for kilonova and afterglow emission, a recurring question is whether there is any detectable \ac{EM} signal in the seconds before, during, and after the merger itself. Since the \ac{GW} inspiral signal is in principal detectable for up to hundreds of seconds before merger, one could imagine positioning rapidly slewing instruments to search for any prompt emission. This concept was explored by the authors \citep{Cannon:2011vi}.

On the topic of very low\nobreakdashes-latency localization, we also recommend Chen~\&~Holz~\cite{ChenHolzRapidLocalization}, who propose a rapid localization scheme that is similar to ours, but even faster because it makes some additional compromises: their likelihood is strictly Gaussian, so one more marginalization integral (the integral over arrival time) can be performed analytically.

\section{Conclusion}

We have presented a novel, fast, accurate, Bayesian algorithm for inferring the sky locations of compact binary merger sources that may soon be detected by advanced ground\nobreakdashes-based \ac{GW} detectors. For \ac{BNS} systems with small spins, we have shown that \ac{BAYESTAR} produces sky maps that are as accurate as the full \ac{MCMC} parameter estimation code but can do so within $\sim$10\,s after a detection. Still faster response times should be possible in the future (if warranted) by deploying \ac{BAYESTAR} on machines with more cores or by distributing \ac{BAYESTAR} across multiple computers.

Following a \ac{BNS} merger, the signal will be detected by the matched\nobreakdashes-filter pipeline within tens of seconds; an alert containing the time and estimated significance of the event can be distributed almost immediately (although a human validation stage that may be present at the beginning of the first observing run may introduce some additional latency). The localization from \ac{BAYESTAR} will be available tens of seconds to a minute later. Finally, the refined localization and the detailed estimates of masses and spins from LALINFERENCE will be distributed hours to days later.

Relevant time scales for possible \ac{EM} counterparts to \ac{GW} signals include seconds (the prompt \ac{GRB} signature), hundreds of seconds (extended emission and X\nobreakdashes-ray plateaus that are observed for some short \acp{GRB}), minutes to hours (X\nobreakdashes-ray and optical afterglow), hours to days (the kilonova or the blue flashes associated with unbound ejecta or disk winds), and days to years (the radio afterglow). For the first time, we are able to provide accurate localizations before the peak of any of these \ac{EM} signatures (except for the short \ac{GRB} or any premerger signal). Even for components like the kilonova that should peak within hours to days, the availability of the localizations within seconds might provide a window of several hours to obtain tiled images of the area before the \ac{EM} emission begins. These could be used as reference images, crucial at optical wavelengths for establishing the rapid rise and quickly distinguishing from slower background transients.

\makeatletter{}\providecommand{\acrolowercase}[1]{\lowercase{#1}}

\begin{acronym}
\acro{2MASS}[2MASS]{Two Micron All Sky Survey}
\acro{AdVirgo}[AdVirgo]{Advanced Virgo}
\acro{AMI}[AMI]{Arcminute Microkelvin Imager}
\acro{AGN}[AGN]{active galactic nucleus}
\acroplural{AGN}[AGN\acrolowercase{s}]{active galactic nuclei}
\acro{aLIGO}[aLIGO]{Advanced \acs{LIGO}}
\acro{ATCA}[ATCA]{Australia Telescope Compact Array}
\acro{ATLAS}[ATLAS]{Asteroid Terrestrial-impact Last Alert System}
\acro{BAT}[BAT]{Burst Alert Telescope\acroextra{ (instrument on \emph{Swift})}}
\acro{BATSE}[BATSE]{Burst and Transient Source Experiment\acroextra{ (instrument on \acs{CGRO})}}
\acro{BAYESTAR}[BAYESTAR]{BAYESian TriAngulation and Rapid localization}
\acro{BBH}[BBH]{binary black hole}
\acro{BHBH}[BHBH]{\acl{BH}\nobreakdashes--\acl{BH}}
\acro{BH}[BH]{black hole}
\acro{BNS}[BNS]{binary neutron star}
\acro{CARMA}[CARMA]{Combined Array for Research in Millimeter\nobreakdashes-wave Astronomy}
\acro{CASA}[CASA]{Common Astronomy Software Applications}
\acro{CFH12k}[CFH12k]{Canada--France--Hawaii $12\,288 \times 8\,192$ pixel CCD mosaic\acroextra{ (instrument formerly on the Canada--France--Hawaii Telescope, now on the \ac{P48})}}
\acro{CRTS}[CRTS]{Catalina Real-time Transient Survey}
\acro{CTIO}[CTIO]{Cerro Tololo Inter-American Observatory}
\acro{CBC}[CBC]{compact binary coalescence}
\acro{CCD}[CCD]{charge coupled device}
\acro{CDF}[CDF]{cumulative distribution function}
\acro{CGRO}[CGRO]{Compton Gamma Ray Observatory}
\acro{CMB}[CMB]{cosmic microwave background}
\acro{CRLB}[CRLB]{Cram\'{e}r\nobreakdashes--Rao lower bound}
\acro{cWB}[\acrolowercase{c}WB]{Coherent WaveBurst}
\acro{DASWG}[DASWG]{Data Analysis Software Working Group}
\acro{DBSP}[DBSP]{Double Spectrograph\acroextra{ (instrument on \acs{P200})}}
\acro{DCT}[DCT]{Discovery Channel Telescope}
\acro{DECAM}[DECam]{Dark Energy Camera\acroextra{ (instrument on the Blanco 4\nobreakdashes-m telescope at \acs{CTIO})}}
\acro{DFT}[DFT]{discrete Fourier transform}
\acro{EM}[EM]{electromagnetic}
\acro{FD}[FD]{frequency domain}
\acro{FAR}[FAR]{false alarm rate}
\acro{FFT}[FFT]{fast Fourier transform}
\acro{FIR}[FIR]{finite impulse response}
\acro{FITS}[FITS]{Flexible Image Transport System}
\acro{FLOPS}[FLOPS]{floating point operations per second}
\acro{FOV}[FOV]{field of view}
\acroplural{FOV}[FOV\acrolowercase{s}]{fields of view}
\acro{FTN}[FTN]{Faulkes Telescope North}
\acro{GBM}[GBM]{Gamma-ray Burst Monitor\acroextra{ (instrument on \emph{Fermi})}}
\acro{GCN}[GCN]{Gamma-ray Coordinates Network}
\acro{GMOS}[GMOS]{Gemini Multi-object Spectrograph\acroextra{ (instrument on the Gemini telescopes)}}
\acro{GRB}[GRB]{gamma-ray burst}
\acro{GSL}[GSL]{GNU Scientific Library}
\acro{GTC}[GTC]{Gran Telescopio Canarias}
\acro{GW}[GW]{gravitational wave}
\acro{HAWC}[HAWC]{High\nobreakdashes-Altitude Water \v{C}erenkov Gamma\nobreakdashes-Ray Observatory}
\acro{HCT}[HCT]{Himalayan Chandra Telescope}
\acro{HEALPix}[HEALP\acrolowercase{ix}]{Hierarchical Equal Area isoLatitude Pixelization}
\acro{HEASARC}[HEASARC]{High Energy Astrophysics Science Archive Research Center}
\acro{HETE}[HETE]{High Energy Transient Explorer}
\acro{HFOSC}[HFOSC]{Himalaya Faint Object Spectrograph and Camera\acroextra{ (instrument on \acs{HCT})}}
\acro{HMXB}[HMXB]{high\nobreakdashes-mass X\nobreakdashes-ray binary}
\acroplural{HMXB}[HMXB\acrolowercase{s}]{high\nobreakdashes-mass X\nobreakdashes-ray binaries}
\acro{HSC}[HSC]{Hyper Suprime\nobreakdashes-Cam\acroextra{ (instrument on the 8.2\nobreakdashes-m Subaru telescope)}}
\acro{IACT}[IACT]{imaging atmospheric \v{C}erenkov telescope}
\acro{IIR}[IIR]{infinite impulse response}
\acro{IMACS}[IMACS]{Inamori-Magellan Areal Camera \& Spectrograph\acroextra{ (instrument on the Magellan Baade telescope)}}
\acro{IPAC}[IPAC]{Infrared Processing and Analysis Center}
\acro{IPN}[IPN]{InterPlanetary Network}
\acro{IPTF}[\acrolowercase{i}PTF]{intermediate \acl{PTF}}
\acro{ISM}[ISM]{interstellar medium}
\acro{KAGRA}[KAGRA]{KAmioka GRAvitational\nobreakdashes-wave observatory}
\acro{KDE}[KDE]{kernel density estimator}
\acro{LAT}[LAT]{Large Area Telescope}
\acro{LCOGT}[LCOGT]{Las Cumbres Observatory Global Telescope}
\acro{LHO}[LHO]{\ac{LIGO} Hanford Observatory}
\acro{LIGO}[LIGO]{Laser Interferometer \acs{GW} Observatory}
\acro{llGRB}[\acrolowercase{ll}GRB]{low\nobreakdashes-luminosity \ac{GRB}}
\acro{LLOID}[LLOID]{Low Latency Online Inspiral Detection}
\acro{LLO}[LLO]{\ac{LIGO} Livingston Observatory}
\acro{LMI}[LMI]{Large Monolithic Imager\acroextra{ (instrument on \ac{DCT})}}
\acro{LOFAR}[LOFAR]{Low Frequency Array}
\acro{LSB}[LSB]{long, soft burst}
\acro{LSC}[LSC]{\acs{LIGO} Scientific Collaboration}
\acro{LSO}[LSO]{last stable orbit}
\acro{LSST}[LSST]{Large Synoptic Survey Telescope}
\acro{LTI}[LTI]{linear time invariant}
\acro{MAP}[MAP]{maximum a posteriori}
\acro{MBTA}[MBTA]{Multi-Band Template Analysis}
\acro{MCMC}[MCMC]{Markov chain Monte Carlo}
\acro{MLE}[MLE]{\ac{ML} estimator}
\acro{ML}[ML]{maximum likelihood}
\acro{NED}[NED]{NASA/IPAC Extragalactic Database}
\acro{NSBH}[NSBH]{neutron star\nobreakdashes--black hole}
\acro{NSBH}[NSBH]{\acl{NS}\nobreakdashes--\acl{BH}}
\acro{NSF}[NSF]{National Science Foundation}
\acro{NSNS}[NSNS]{\acl{NS}\nobreakdashes--\acl{NS}}
\acro{NS}[NS]{neutron star}
\acro{OT}[OT]{optical transient}
\acro{P48}[P48]{Palomar 48~inch Oschin telescope}
\acro{P60}[P60]{robotic Palomar 60~inch telescope}
\acro{P200}[P200]{Palomar 200~inch Hale telescope}
\acro{PC}[PC]{photon counting}
\acro{PSD}[PSD]{power spectral density}
\acro{PSF}[PSF]{point-spread function}
\acro{PTF}[PTF]{Palomar Transient Factory}
\acro{QUEST}[QUEST]{Quasar Equatorial Survey Team}
\acro{RAPTOR}[RAPTOR]{Rapid Telescopes for Optical Response}
\acro{REU}[REU]{Research Experiences for Undergraduates}
\acro{RMS}[RMS]{root mean square}
\acro{ROTSE}[ROTSE]{Robotic Optical Transient Search}
\acro{S5}[S5]{\ac{LIGO}'s fifth science run}
\acro{S6}[S6]{\ac{LIGO}'s sixth science run}
\acro{SAA}[SAA]{South Atlantic Anomaly}
\acro{SHB}[SHB]{short, hard burst}
\acro{SHGRB}[SHGRB]{short, hard \acl{GRB}}
\acro{SMT}[SMT]{Slewing Mirror Telescope\acroextra{ (instrument on \acs{UFFO} Pathfinder)}}
\acro{SNR}[S/N]{signal\nobreakdashes-to\nobreakdashes-noise ratio}
\acro{SSC}[SSC]{synchrotron self\nobreakdashes-Compton}
\acro{SDSS}[SDSS]{Sloan Digital Sky Survey}
\acro{SED}[SED]{spectral energy distribution}
\acro{SN}[SN]{supernova}
\acroplural{SN}[SN\acrolowercase{e}]{supernova}
\acro{SNIcBL}[\acs{SN}\,I\acrolowercase{c}\nobreakdashes-BL]{broad\nobreakdashes-line Type~Ic \ac{SN}}
\acroplural{SNIcBL}[\acsp{SN}\,I\acrolowercase{c}\nobreakdashes-BL]{broad\nobreakdashes-line Type~Ic \acp{SN}}
\acro{SVD}[SVD]{singular value decomposition}
\acro{TAROT}[TAROT]{T\'{e}lescopes \`{a} Action Rapide pour les Objets Transitoires}
\acro{TDOA}[TDOA]{time delay on arrival}
\acroplural{TDOA}[TDOA\acrolowercase{s}]{time delays on arrival}
\acro{TD}[TD]{time domain}
\acro{TOA}[TOA]{time of arrival}
\acroplural{TOA}[TOA\acrolowercase{s}]{times of arrival}
\acro{TOO}[TOO]{target\nobreakdashes-of\nobreakdashes-opportunity}
\acroplural{TOO}[TOO\acrolowercase{s}]{targets of opportunity}
\acro{UFFO}[UFFO]{Ultra Fast Flash Observatory}
\acro{UHE}[UHE]{ultra high energy}
\acro{UVOT}[UVOT]{UV/Optical Telescope\acroextra{ (instrument on \emph{Swift})}}
\acro{VHE}[VHE]{very high energy}
\acro{VLA}[VLA]{Karl G. Jansky Very Large Array}
\acro{VLT}[VLT]{Very Large Telescope}
\acro{WAM}[WAM]{Wide\nobreakdashes-band All\nobreakdashes-sky Monitor\acroextra{ (instrument on \emph{Suzaku})}}
\acro{WCS}[WCS]{World Coordinate System}
\acro{WSS}[w.s.s.]{wide\nobreakdashes-sense stationary}
\acro{XRF}[XRF]{X\nobreakdashes-ray flash}
\acroplural{XRF}[XRF\acrolowercase{s}]{X\nobreakdashes-ray flashes}
\acro{XRT}[XRT]{X\nobreakdashes-ray Telescope\acroextra{ (instrument on \emph{Swift})}}
\acro{ZTF}[ZTF]{Zwicky Transient Facility}
\end{acronym}

\begin{acknowledgements}
This paper has LIGO Document Number LIGO\nobreakdashes-P1500009\nobreakdashes-v8.

We thank John Veitch and Will Farr for chairing a review of the analysis and code. We thank Britt Griswold for assistance with preparing Fig.~\ref{fig:adaptive-illustration}.
\end{acknowledgements}

LIGO was constructed by the California Institute of Technology and Massachusetts Institute of Technology with funding from the \ac{NSF} and operates under Cooperative Agreement No.~PHY\nobreakdashes-0107417.

This research was supported by the \ac{NSF} through a Graduate Research Fellowship to L.~S. L.~S. thanks the Aspen Center for Physics and \ac{NSF} Grant \#1066293 for hospitality during the editing of this paper.

Source code for \ac{BAYESTAR} is available as part of LALInference,\footnote{\url{https://ligo-vcs.phys.uwm.edu/cgit/lalsuite/tree/lalinference}.} the open source LIGO/Virgo parameter estimation toolchain, which is in turn part of LALSuite.\footnote{\url{http://www.lsc-group.phys.uwm.edu/daswg/projects/lalsuite.html}.} This research made use of Astropy\footnote{\url{http://www.astropy.org}.}~\citep{astropy}, a community-developed core Python package for astronomy. Some of the results in this paper have been derived using \ac{HEALPix}\footnote{\url{http://healpix.sourceforge.net}.}~\cite{healpix}.

Some results were produced on the NEMO computing cluster operated by the Center for Gravitation and Cosmology at University of Wisconsin\nobreakdashes--Milwaukee under \ac{NSF} Grants No.~PHY\nobreakdashes-0923409 and No.~PHY\nobreakdashes-0600953.

\appendix

\section{Independence of intrinsic and extrinsic errors}
\label{sec:independence-intrinsic-extrinsic}

If all of the detectors have the same noise \acp{PSD} up to multiplicative factors, $c_1 S_1(\omega) = c_2 S_2(\omega) = \cdots = c_n S_n(\omega) \equiv S(\omega)$, then we can show that the errors in the intrinsic parameters (masses) are not correlated with sky position errors. This is because we can change variables from amplitudes, phases, and times to amplitude ratios, phase differences, and time differences. With $N$ detectors, we can form a single average amplitude, time, and phase, plus $N-1$ linearly independent differences. The averages are correlated with the intrinsic parameters, but neither are correlated with the differences. Since only the differences inform the sky location, this gives us license to neglect uncertainty in masses when we are computing the sky resolution.

This is easiest to see if we make the temporary change of variables $\rho \rightarrow \varsigma = \log \rho$. This allows us to factor out the \ac{SNR} dependence from the single-detector Fisher matrix. The extrinsic part becomes
\begin{align}
    \mathcal{I}_{\bm\theta_i,\bm\theta_i} &= \bordermatrix{
        ~ & \varsigma_i & \gamma_i & \tau_i \cr
        \varsigma_i & {\rho_i}^2 & 0 & 0 \cr
        \gamma_i & 0 & {\rho_i}^2 & -{\rho_i}^2 {\overline{\omega}}_i \cr
        \tau_i & 0 & -{\rho_i}^2 {\overline{\omega}}_i & {\rho_i}^2 {\overline{\omega^2}}_i
    }
    \\
    \nonumber
    &= {\rho_i}^2 \left( \begin{array}{ccc}
        1 & 0 & 0 \\
        0 & 1 & -{\overline{\omega}}_i \\
        0 & -{\overline{\omega}}_i & {\overline{\omega^2}}_i
    \end{array} \right).
\end{align}
Due to our assumption that the detectors' \acp{PSD} are proportional to each other, the noise moments are the same for all detectors, ${\overline{\omega^k}}_i \equiv {\overline{\omega^k}}$. Then, we can write the single-detector Fisher matrix as
\begin{equation}
    \mathcal{I}_i = {\rho_i}^2 \left(
        \begin{array}{cc}
            A & B \\
            B\transpose & C
        \end{array}
    \right),
\end{equation}
with the top-left block $A$ comprising the extrinsic parameters and the bottom-right block $C$ the intrinsic parameters.

Information is additive, so the Fisher matrix for the whole detector network is
\begin{equation}
    \mathcal{I}_\mathrm{net} = \left(
    \begin{array}{ccccc}
    {\rho_1}^2 A & 0 & \cdots & 0 & {\rho_1}^2 B \\
    0 & {\rho_2}^2 A & & \vdots & {\rho_1}^2 B \\
    \vdots & & \ddots & 0 & \vdots \\
    0 & 0 & \cdots & {\rho_N}^2 A & {\rho_N}^2 B \\
    {\rho_1}^2 B\transpose & {\rho_2}^2 B\transpose & \cdots & {\rho_N}^2 B\transpose & {\rho_\mathrm{net}}^2 C
    \end{array}
    \right).
\end{equation}
Now we introduce the change of variables that sacrifices the $N$th detector's extrinsic parameters for the network averages,
\begin{align}
    &\begin{array}{c@{\quad\rightarrow\quad}l}
    \varsigma_N & \overline{\varsigma} = \left(\sum_i {\rho_i}^2 \varsigma_i\right) / {\rho_\mathrm{net}}^2, \\
    \gamma_N & \overline{\gamma} = \left(\sum_i {\rho_i}^2 \gamma_i\right) / {\rho_\mathrm{net}}^2, \\
    \tau_N & \overline{\tau} = \left(\sum_i {\rho_i}^2 \tau_i\right) / {\rho_\mathrm{net}}^2, \\
    \end{array}
\intertext{and replaces the first $N-1$ detectors' extrinsic parameters with differences,}
    &\left.
    \begin{array}{c@{\quad\rightarrow\quad}l}
    \varsigma_i & \delta\varsigma_i = \varsigma_i - \overline{\varsigma} \\
    \gamma_i & \delta\gamma_i = \gamma_i - \overline{\gamma} \\
    \tau_i & \delta\tau_i = \tau_i - \overline{\tau} \\
    \end{array}
    \right\} \text{ for } i = 1, \dots, N - 1.
\end{align}
The Jacobian matrix that describes this change of variables is
\begin{equation}
    J = \left(
    \begin{array}{cccccc}
    1 & 0 & \cdots & 0 & 1 & 0 \\
    0 & 1 & & 0 & 1 & 0 \\
    \vdots & & \ddots & & \vdots & \vdots \\
    0 & 0 & \cdots & 1 & 1 & 0 \\
    \frac{-{\rho_1}^2}{{\rho_N}^2} & \frac{-{\rho_2}^2}{{\rho_N}^2} & \cdots & \frac{-{\rho_{N-1}}^2}{{\rho_N}^2} & 1 & 0 \\
    0 & 0 & \cdots & 0 & 0 & 1 \\
    \end{array}
    \right).
\end{equation}
The transformed network Fisher matrix is block diagonal,
\begin{widetext}
\begin{equation}
    \mathcal{I}_\mathrm{net} \rightarrow J\transpose \mathcal{I}_\mathrm{net} J = \left(
    \begin{array}{cccccc}
    {\rho_1}^2(1+\frac{1}{{\rho_1}^4}) A & \frac{{\rho_1}^2 {\rho_2}^2}{{\rho_N}^2} A & \cdots & \frac{{\rho_1}^2 {\rho_{N-1}}^2}{{\rho_N}^2} A & 0 & 0 \\
    \frac{{\rho_1}^2 {\rho_2}^2}{{\rho_N}^2} A & {\rho_2}^2(1+\frac{1}{{\rho_1}^4}) A & & \frac{{\rho_2}^2 {\rho_{N-1}}^2}{{\rho_N}^2} A & 0 & 0 \\
    \vdots & & \ddots & \vdots & \vdots & \vdots \\
    \frac{{\rho_1}^2 {\rho_{N-1}}^2}{{\rho_N}^2} A & \frac{{\rho_2}^2 {\rho_{N-1}}^2}{{\rho_N}^2} A & \cdots & {\rho_{N-1}}^2(1+\frac{1}{{\rho_1}^4}) A & 0 & 0 \\
    0 & 0 & \cdots & 0 & {\rho_\mathrm{net}}^2 A & {\rho_\mathrm{net}}^2 B \\
    0 & 0 & \cdots & 0 & {\rho_\mathrm{net}}^2 B\transpose & {\rho_\mathrm{net}}^2 C \\
    \end{array}
    \right).
\end{equation}
\end{widetext}
The top\nobreakdashes-left block contains $N-1$ relative amplitudes, phases, and times on arrival, all potentially correlated with each other. The bottom\nobreakdashes-right block contains the average amplitudes, phases, and times, as well as the masses. The averages and the masses are correlated with each other, but are not correlated with the differences. Because only the differences are informative for sky localization, we drop the intrinsic parameters from the rest of the Fisher matrix calculations in the Appendix.

\section{Interpretation of phase and time errors}
\label{sec:interpretation-of-errors}

The Fisher matrix in Eq.~(\ref{eq:fisher-matrix}) is block diagonal, which implies that estimation errors in the signal amplitude $\rho$ are uncorrelated with the phase $\gamma$ and time $\tau$. A sequence of two changes of variables lends some physical interpretation to the nature of the coupled estimation errors in $\gamma$~and~$\tau$.

First, we put the phase and time on the same footing by measuring the time in units of $1 / \sqrt{\overline{\omega^2}}$ with a change of variables from $\tau$ to $\gamma_\tau = \sqrt{\overline{\omega^2}} \tau$:
\begin{equation}
    \mathcal{I}' = \bordermatrix{
        ~ & \rho_i & \gamma_i & \gamma_{\tau,i} \cr
        \rho_i & 1 & 0 & 0 \cr
        \gamma_i & 0 & {\rho_i}^2 & -{\rho_i}^2\frac{{\overline{\omega}}_i}{\sqrt{{\overline{\omega^2}}_i}} \cr
        \gamma_{\tau,i} & 0 & -{\rho_i}^2\frac{{\overline{\omega}}_i}{\sqrt{{\overline{\omega^2}}_i}} & {\rho_i}^2
    }.
\end{equation}
The second change of variables, from $\gamma$ and $\gamma_\tau$ to $\gamma_\pm = \frac{1}{\sqrt{2}}(\gamma \pm \gamma_\tau)$, diagonalizes the Fisher matrix:
\begin{equation}\label{eq:fisher-matrix-extrinsic-one-detector}
    \mathcal{I}'' = \bordermatrix{
        ~ & \rho_i & \gamma_{+,i} & \gamma_{-,i} \cr
        \rho_i & 1 & 0 & 0 \cr
        \gamma_{+,i} & 0 & \left(1 - \frac{\overline{\omega}_i}{\sqrt{\overline{\omega^2}_i}}\right){\rho_i}^2 & 0 \cr
        \gamma_{-,i} & 0 & 0 & \left(1 + \frac{\overline{\omega}_i}{\sqrt{\overline{\omega^2}_i}}\right){\rho_i}^2
    }.
\end{equation}
Thus, in the appropriate time units, the \textit{sum and difference} of the phase and time of the signal are measured independently.

\section{Position resolution}
\label{sec:position-resolution}

Finally, we will calculate the position resolution of a network of \ac{GW} detectors. We could launch directly into computing derivatives of the full signal model from Eq.~(\ref{eq:full-signal-model}) with respect to all of the parameters, but this would result in a very complicated expression. Fortunately, we can take two shortcuts. First, since we showed in Appendix~\ref{sec:independence-intrinsic-extrinsic} that the intrinsic parameters are correlated only with an overall nuisance average arrival time, amplitude, and phase, we need not consider the derivatives with respect to mass at all. Second, we can reuse the extrinsic part of the single-detector Fisher matrix from Eq.~(\ref{eq:fisher-matrix}) by computing the much simpler Jacobian matrix to transform from the time, amplitude, and phase on arrival to the parameters of interest.

We begin by transforming the single\nobreakdashes-detector Fisher matrix from a polar to a rectangular representation of the complex amplitude given in Eqs.~(\ref{eq:optimal-rho}, \ref{eq:optimal-tau}), $\rho_i, \gamma_i \rightarrow \Re[z_i] = \rho_i \cos \gamma_i, \Im[z_i] = \rho_i \sin \gamma_i$:
\begin{equation}\label{eq:fisher-matrix-extrinsic-one-detector-cartesian}
    \mathcal{I}_i = \bordermatrix{
        ~ & \Re[z_i] & \Im[z_i] & \tau_i \cr
                \Re[z_i] &
        1 &
        0 &
        {\overline{\omega}}_i b_i \cr
                \Im[z_i] &
        0 &
        1 &
        -{\overline{\omega}}_i b_i \cr
                \tau_i &
        {\overline{\omega}}_i b_i &
        -{\overline{\omega}}_i b_i &
        {\rho_i}^2 {\overline{\omega^2}}_i
    }.
\end{equation}

Consider a source in a ``standard'' orientation with the direction of propagation along the $+z$ axis, such that the \ac{GW} polarization tensor may be written in Cartesian coordinates as
\begin{equation}
    H = \frac{1}{r} e^{2 i \phi_c} \left(
    \begin{array}{ccc}
        \frac{1}{2}(1 + \cos^2 \iota) & i \cos\iota & 0 \\
        i \cos\iota & -\frac{1}{2}(1 + \cos^2 \iota) & 0 \\
        0 & 0 & 0
    \end{array}
    \right).
\end{equation}
Now introduce a rotation matrix $R$ that actively transforms this source to the Earth-relative polar coordinates $\theta, \phi$, and gives the source a polarization angle $\psi$ (adopting temporarily the notation $\mathrm{c}_\theta = \cos\theta,\,\mathrm{s}_\theta = \sin\theta$):
\begin{widetext}
\begin{equation}
    R = R_z(\phi) R_y(\theta) R_z(\psi) R_y(\pi)
    = \left(
        \begin{array}{ccc}
            \mathrm{c}_\phi & -\mathrm{s}_\phi & 0 \\
            \mathrm{s}_\phi & -\mathrm{c}_\phi & 0 \\
            0 & 0 & 1
        \end{array}
    \right)
    \left(
        \begin{array}{ccc}
            \mathrm{c}_\theta & 0 & \mathrm{s}_\theta \\
            0 & 1 & 0 \\
            -\mathrm{s}_\theta & 0 & \mathrm{c}_\theta
        \end{array}
    \right)
    \left(
        \begin{array}{ccc}
            \mathrm{c}_\psi & -\mathrm{s}_\psi & 0 \\
            \mathrm{s}_\psi & -\mathrm{c}_\psi & 0 \\
            0 & 0 & 1
        \end{array}
    \right)
    \left(
        \begin{array}{ccc}
            -1 & 0 & 0 \\
            0 & 1 & 0 \\
            0 & 0 & -1
        \end{array}
    \right).
\end{equation}
\end{widetext}
(The rightmost rotation reverses the propagation direction so that the wave is traveling \emph{from} the sky position $\theta, \phi$.) With the (symmetric) detector response tensor $D_i$, we can write the received amplitude and arrival time as
\begin{align}
    z_i &= r_{1,i} \Tr \left[ D_i R \, H \, R\transpose \right], \\
    \tau_i &= t_\oplus + {\mathbf{d}_i}\transpose R \, \mathbf{k}.
\end{align}
Equivalently, we can absorb the rotation $R$ and the horizon distance $r_{1,i}$ into the polarization tensor, detector response tensors, and positions,
\begin{align}
    H &\rightarrow H^\prime = R_z(\psi) \, R_y(\pi) \, H \, R_y(\pi)\transpose  \, R_z(\psi)\transpose, \\
    D_i &\rightarrow D_i^\prime = r_{1,i} \, R_y(\theta)\transpose \, R_z(\phi)\transpose \, D_i \, R_z(\phi) \, R_y(\theta), \\
    \mathbf{d}_i &\rightarrow \mathbf{d}_i^\prime = R_y(\theta)\transpose \, R_z(\phi)\transpose \, \mathbf{d}_i, \\
    \mathbf{k} &\rightarrow \mathbf{k}^\prime = (0, 0, -1).
\end{align}
Now the model becomes
\begin{align}
    \label{eq:H-hplus-hcross}
    H^\prime &= \left(
        \begin{array}{ccc}
            h_+ & h_\times & 0 \\
            h_\times & -h_+ & 0 \\
            0 & 0 & 0
        \end{array}
    \right), \\
    \label{eq:z-hplus-hcross}
    z_i &= \Tr \left[ D_i^\prime H^\prime \right] = h_+ (D_{00}^\prime - D_{11}^\prime) + 2 h_\times D_{01}^\prime, \\
    \tau_i &= t_\oplus + (\mathbf{d}_i^\prime) \cdot \mathbf{k}, \\
        \intertext{where}
        \label{eq:hplus}
    h_+ &= \frac{1}{r} e^{2 i \phi_c} \left[\frac{1}{2} (1+\cos^2\iota) \cos{2\psi} + i \cos\iota \sin{2\psi}\right], \\
    \label{eq:hcross}
    h_\times &= \frac{1}{r} e^{2 i \phi_c} \left[\frac{1}{2} (1+\cos^2\iota) \sin{2\psi} - i \cos\iota \cos{2\psi}\right].
\end{align}
We insert an infinitesimal rotation $\delta R$ to perturb the source's orientation from the true value:
\begin{align}
    z_i &= \Tr \left[ D_i^\prime (\delta R) H^\prime (\delta R)\transpose \right], \\
    \tau_i &= t_\oplus + (\mathbf{d}_i^\prime)\transpose (\delta R) \mathbf{k}^\prime.
\end{align}
We only need a first-order expression for $\delta R$, because we will be taking products of first derivatives of it\footnote{Caution: the angles $\delta\theta$ and $\delta\phi$ represent displacements in two orthogonal directions, but are \emph{not} necessarily simply related to $\theta$ and $\phi$.}:
\begin{equation}
    \delta R = \left(
        \begin{array}{ccc}
            1 & 0 & \delta\theta \\
            0 & 1 & \delta\phi \\
            -\delta\theta & -\delta\phi & 1
        \end{array}
    \right).
\end{equation}
We construct a Jacobian matrix $J_i$ to transform from the single\nobreakdashes-detector observables $(\Re[z_i], \Im[z_i], \tau_i)$ to the position perturbations, polarization components, and geocentered arrival time \\ $(\delta\theta, \delta\phi, \Re[h_+], \Im[h_+], \Re[h_\times], \Im[h_\times], t_\oplus)$:
\begin{widetext}
\begin{equation}\label{eq:jacobian-detector-tensor}
    {J_i}\transpose = \bordermatrix{
        ~ & \Re[z_i] & \Im[z_i] & \tau_i \cr
                \delta\theta &
        -2\Re[h_+]D_{02}^\prime -2\Re[h_\times]D_{12}^\prime &
        -2\Im[h_+]D_{02}^\prime -2\Im[h_\times]D_{12}^\prime &
        -d_0^\prime \cr
                \delta\phi &
        -2\Re[h_\times]D_{02}^\prime +2\Re[h_+]D_{12}^\prime &
        -2\Im[h_\times]D_{02}^\prime +2\Im[h_+]D_{12}^\prime &
        -d_1^\prime \cr
                \Re[h_+] & D_{00}^\prime - D_{11}^\prime & 0 & 0 \cr
                \Im[h_+] & 0 & D_{00}^\prime - D_{11}^\prime & 0 \cr
                \Re[h_\times] & 2D_{01}^\prime & 0 & 0 \cr
                \Im[h_\times] & 0 & 2D_{01}^\prime & 0 \cr
                t_\oplus & 0 & 0 & 1
    }.
\end{equation}
\end{widetext}
We transform and sum the information from each detector,
\begin{equation}\label{eq:summed-information}
    \mathcal{I}_\mathrm{net} = \sum_i {J_i}\transpose \mathcal{I}_i {J_i}.
\end{equation}

\subsection{Marginalization over nuisance parameters}

To extract an area from the Fisher matrix, we must first marginalize or discard the nuisance parameters. Note that marginalizing parameters of a multivariate Gaussian distribution amounts to simply dropping the relevant entries in the mean vector and covariance matrix. Since the information is the inverse of the covariance matrix, we need to invert the Fisher matrix, drop all but the first two rows and columns, and then invert again.

This procedure has a shortcut called the Schur complement (see, for example, \citeauthor{numerical-recipes-inversion-partition}~\cite{numerical-recipes-inversion-partition}). Consider a partitioned square matrix $M$ and its inverse:
\begin{equation}
    M = \left(
        \begin{array}{cc}
            A & B \\
            C & D
        \end{array}
    \right), \qquad
    M^{-1} = \left(
        \begin{array}{cc}
            \tilde{A} & \tilde{B} \\
            \tilde{C} & \tilde{D}
        \end{array}
    \right).
\end{equation}
If $A$ and $B$ are square matrices, then the upper-left block of the inverse can be written as
\begin{equation}
    \tilde{A}^{-1} = A - B D^{-1} C.
\end{equation}
If we partition the $\mathcal{I}_\mathrm{net}$ similarly, the $A$ block consists of the first two rows and columns and $D$ is the lower right block that describes all other parameters. Because the Fisher matrix is symmetric, the off\nobreakdashes-diagonal blocks satisfy $C = B\transpose$. Then the Schur complement
\begin{equation}\label{eq:marginal-fisher-matrix}
    \mathcal{I}_\mathrm{marg} = A - B D^{-1} B\transpose
\end{equation}
gives us the information matrix marginalized over all parameters but $\delta\theta$ and $\delta\phi$.

\subsection{Spatial interpretation}

How do we extract the dimensions of the localization from the Fisher matrix? If there are $N \leq 2$ detectors, then the Fisher matrix must be degenerate, because there are $3N$ measurements and seven parameters:
\begin{align*}
    \left\{
    \begin{array}{c}
        \delta\theta \\
        \delta\phi \\
        \Re[h_+] \\
        \Im[h_+] \\
        \Re[h_\times] \\
        \Im[h_\times] \\
        t_\oplus
    \end{array}
    \right\}
    =& \; 7 \text{ parameters} \;
    \longleftrightarrow
    \left\{
    \begin{array}{c}
        \Re[z_i] \\
        \Im[z_i] \\
        \tau_i
    \end{array}
    \right\} \\
    &\times N = 3N \text{ observables.}
\end{align*}
Therefore, for $N=2$ detectors, the marginalized Fisher matrix $\mathcal{I}_\mathrm{marg}$ is singular. Its only nonzero eigenvalue $\lambda$ describes the width of an annulus on the sky. The width of the annulus that contains probability $p$ is given by
\begin{equation}\label{eq:l-pth-quantile}
    L_p = 2 \sqrt{2} \erf^{-1}(p) / \sqrt{\lambda}.
\end{equation}
The prefactor $2 \sqrt{2} / \erf^{-1}(p)$ is the central interval of a normal distribution that contains a probability $p$, and is $\approx 3.3$ for $p = 0.9$. \emph{Caution:} for two\nobreakdashes-detector networks, priors play an important role in practical parameter estimation and areas can be much smaller than one would predict from the Fisher matrix.

For $N \geq 3$ detectors, the parameters are overconstrained by the data, and the Fisher matrix describes the dimensions of an ellipse. Within a circle of radius $r$ centered on the origin, the enclosed probability $p$ is
\begin{equation}
    p = \int_0^{2\pi} \int_0^r \frac{1}{2\pi} e^{-s^2/2} s \, ds \, d\phi = 1 - e^{-r^2/2}.
\end{equation}
Therefore the radius $r$ of the circle that contains a probability $p$ is
\begin{equation}
    r = \sqrt{-2\ln (1 - p)}.
\end{equation}
Suppose that the eigenvalues of the Fisher matrix are $\lambda_1$ and $\lambda_2$. This describes a $1\sigma$ uncertainty ellipse that has major and minor radii ${\lambda_1}^{-1/2}$, ${\lambda_2}^{-1/2}$, and area $A_{1\sigma} = \pi / \sqrt{\lambda_1 \lambda_2} = \pi / \sqrt{\det\mathcal{I}}$. Then, the area of an ellipse containing probability $p$ is
\begin{equation}\label{eq:a-pth-quantile}
    A_p = -2\pi\ln(1-p) / \sqrt{\det \mathcal{I}},
\end{equation}
or, more memorably for the $90$th percentile, $A_{0.9} = 2\pi\ln(10) / \sqrt{\det{\mathcal{I}}}$.

\subsection{Outline of calculation}
\label{sec:fisher-matrix-area-outline}

Using the above derivation, we arrive at a prediction for the sky resolution of a \ac{GW} detector network. We took some shortcuts that allowed us to avoid directly evaluating the complicated derivatives of the signal itself with respect to the sky location. As a result, the expressions involved in each step are simple enough to be manually entered into a computer program. However, because the procedure involves several steps, we outline it once again below.

\begin{enumerate}
    \item Compute, for each detector, the horizon distance $r_{1,i}$, the angular frequency moments ${\overline{\omega}}_i$ and ${\overline{\omega^2}}_i$, and $(h_+, h_\times)$ from Eqs.~(\ref{eq:hplus},~\ref{eq:hcross}). (These can be reused for multiple source positions as long as the masses and the detector noise \acp{PSD} are the same.)
    \item For a given $\phi, \theta, \psi$, compute the complex received amplitude $z_i$ from Eqs.~(\ref{eq:H-hplus-hcross},~\ref{eq:z-hplus-hcross}), the extrinsic Fisher matrix from Eq.~(\ref{eq:fisher-matrix}), and the Jacobian from Eq.~(\ref{eq:jacobian-detector-tensor}).
    \item Sum the information from all detectors using Eq.~\ref{eq:summed-information}.
    \item Compute the marginalized Fisher matrix from the Schur complement using Eq.~(\ref{eq:marginal-fisher-matrix}).
    \item If there are two detectors, find the width $L_p$ of the ring describing the $p$th quantile using Eq.~(\ref{eq:l-pth-quantile}). If there are three or more detectors, find the area $A_p$ of the $p$th quantile using Eq.~(\ref{eq:a-pth-quantile}).
    \item {[}Optionally, convert from (ste)radians to (square) degrees.{]}
\end{enumerate}

See the code listing in Appendix~A.6 of Ref.~\cite{leo-singer-thesis}.

\bibliography{apj-jour,bib/telescope,ms}

\end{document}